\documentclass[a4paper,10pt,onecolumn]{article}
\usepackage[utf8]{inputenc}
\usepackage{amsmath,amssymb,amsfonts,cite}
\usepackage[]{fontenc}
\usepackage[scaled]{helvet}
\usepackage{pdflscape}
\usepackage{graphicx}
\usepackage{pict2e}
\usepackage{enumerate}
\usepackage{setspace}
\usepackage{xcolor}
\usepackage{multirow}
\usepackage{amsbsy} % boldsymbols
\usepackage{bm}
\usepackage{gensymb}
\usepackage{boxedminipage}
\usepackage{framed}
\usepackage[font=small]{caption}
\usepackage[affil-it]{authblk} 
\usepackage{hyperref}
\usepackage{etoolbox}
\usepackage{lmodern}
\usepackage{booktabs}
\usepackage{tabularx}
\usepackage{siunitx}
\usepackage{array}
\usepackage{indentfirst}
\usepackage{enumitem}
\usepackage{wasysym} 
\usepackage{makecell}
\usepackage{microtype}
\usepackage{natbib}
\usepackage{chngcntr}
\usepackage{verbatim}
\usepackage{adjustbox}
\DeclareCaptionFormat{extended}{Extended Figure \thefigure: #1#2#3}

\begin{document}
\title{The Morpho-Kinetic Landscape of Macrophage Modes During Wound Healing in Zebrafish }

\author[1,*]{Seol Ah Park}
\author[1]{Giulia Lupi}
\author[2]{Resul Ozbilgic}
\author[1]{Michal Koll\'ar}
\author[1]{Aneta A. O\v{z}vat}
% \author[2]{Georges Lutfalla}
\author[2]{Mai Nguyen-Chi}
\author[1]{Karol Mikula}

\affil[1]{\footnotesize Department of Mathematics and Descriptive Geometry, Faculty of Civil Engineering, Slovak University of Technology, Slovak Republic}
\affil[2]{\footnotesize LPHI, Université de Montpellier, CNRS, INSERM, Montpellier, France}

\affil[*]{seol.park@stuba.sk}

%\affil[+]{these authors contributed equally to this work}

%\keywords{Keyword1, Keyword2, Keyword3}
\date{}
\maketitle
\begin{abstract}
Macrophages play an essential role in wound healing due to their dynamic nature and functional plasticity, exhibiting highly heterogeneous morpho-kinetic behaviors depending on their activation states. However, quantitative analysis of macrophage behavior in \textit{in vivo} settings remains limited, largely due to the complexity of their diverse morphologies and motility patterns over time.
In this study, we present an analytic workflow to investigate macrophage dynamics in zebrafish.  By computing a comprehensive set of morpho-kinetic features, we observe that M1 (pro-inflammatory) and M2 (anti-inflammatory) macrophages exhibit distinct behaviors, such as reduced shape elongation, more directed movement, and less random-like motion in M1 compared to M2 macrophages.
Based on these features, we classify macrophages in the transition period into M1-like and M2-like groups.
We compare and analyze their behaviors, which allows us to estimate the timing of the phenotypic switch. In addition, macrophages not expressing Tumor Necrosis Factor (TNF) are located significantly farther from the wound compared to M1 macrophages. While macrophages unstimulated by wound signaling exhibit some M2-like features, they differ notably in shape elongation and migration speed.
In summary, this study provides a quantitative analysis of macrophage behavior during wound healing and suggests distinct behavioral landscapes across different macrophage activation states. 
\end{abstract}
%\begin{document}

%Wound healing is an essential biological process that restores tissue integrity and protects against pathogens.
%differ not only in the aforementioned traits but also in their distance from the wound and changes in speed over time. Our analysis further suggests that the phenotypic switch from M1 to M2 likely may occur approximately between 460 and 550 minutes post-amputation. 

%\flushbottom
\maketitle
\thispagestyle{empty}

%\noindent Please note: Abbreviations should be introduced at the first mention in the main text – no abbreviations lists. Suggested structure of main text (not enforced) is provided below.

\section*{Introduction}

Wound healing is a fundamental biological process that restores tissue integrity following injury and provides protection against pathogens. Efficient healing is therefore fundamental for maintaining overall health, preventing infections, and reducing healthcare burdens \cite{sonnemann2011wound}.
Among the various immune cells involved in wound healing, macrophages are essential to wound healing, given their dynamic nature and functional plasticity.
As key components of the innate immune system, macrophages take on the role of frontline defenders against pathogens and contribute to tissue homeostasis, repair, and the progression of various diseases, including cancer, chronic inflammation, and infections \cite{wynn2013macrophage}. 

Traditionally, macrophages in wound healing have been described in two main phases \cite{hesketh2017macrophage}: the early recruitment of pro-inflammatory M1 macrophages, followed by a transition to anti-inflammatory M2 macrophages. M1 macrophages dominate the initial inflammatory phase, producing cytokines such as Tumor Necrosis Factor (TNF) and Interleukin-6 (IL-6), and exhibiting strong phagocytic activity to clear pathogens and debris \cite{benoit2008macrophage, daley2010phenotype}. As inflammation subsides, macrophages shift toward the M2 phenotype, supporting tissue repair and resolution through factors like Transforming Growth Factor-$\beta$1 (TGF-$\beta$1) \cite{lucas2010differential, koh2011inflammation}. It has been observed that their polarization state correlates with distinct morphological and migratory features \cite{friedl2008interstitial, van2010matrix, barros2017live, cui2018distinct}. M1 macrophages are typically rounded and flattened, while M2 macrophages are elongated. Correspondingly, M1 cells predominantly use fast, adhesion-independent amoeboid migration, whereas M2 cells exhibit slower, adhesion-dependent mesenchymal movement \cite{adebowale2025monocytes}.
However, it has been reported that macrophage activation occurs along a continuum, including intermediate and non-classical phenotypes, shaped by local environmental cues and not strictly conforming to the M1/M2 framework \cite{mosser2008exploring, strizova2023m1, ghamangiz2025reprogram}. A comprehensive understanding of this spectrum of macrophage modes is therefore crucial, as it provides valuable insights into macrophage functions during tissue repair and helps guide therapeutic strategies for inflammatory and wound healing disorders.

\textit{In vivo} experiments provide unique opportunities to observe the temporal and spatial dynamics of macrophage plasticity and responsiveness during wound healing, offering valuable insights into their functional dynamics and regulatory mechanisms. While several studies have quantitatively analyzed macrophage dynamics in \textit{in vivo} settings \cite{nguyen2015identification, miskolci2019distinct, sipka2022macrophages}, detailed quantitative analysis of the relationship between macrophage activation states and morpho-kinetic changes remains challenging. A major obstacle is the complexity inherent in analyzing macrophages in dynamic \textit{in vivo} environments, where cells exhibit heterogeneous morphologies and diverse motility patterns over time.

In this paper, we investigate various aspects of macrophage behavior \textit{in vivo} using the zebrafish model during wound healing. We examine not only the classical M1 and M2 phenotypes but also macrophages that do not express TNF (Non-M1), unstimulated macrophages (M0), and macrophages in the transitional phase from M1 to M2.
To understand the landscape of macrophage behavior, we analyze it from a morpho-kinetic perspective that captures both morphological and dynamic features. This approach has been demonstrated to be effective for characterizing and clustering the behavior of neutrophils and dendritic cells under inflammatory conditions \cite{crainiciuc2022behavioural}.
Each macrophage mode is quantitatively characterized through detailed morpho-kinetic analysis using advanced image processing techniques tailored to capture complex cellular dynamics.
These methods include macrophage segmentation and tracking \cite{park2023segmentation}, as well as the extraction of directional versus random-like motion \cite{lupi2025mathematical}. 
From time-lapse imaging data, we extract morpho-kinetic features that distinguish macrophage behaviors across modes. Additionally, we apply classification methods \cite{mikula2023natural, mikula2023natural2} to identify M1 and M2 phenotypes specifically during the transition period. This comprehensive analysis reveals key morpho-kinetic features associated with each macrophage mode and elucidates their relationships. We further identify distinguishing characteristics of M1 and M2 macrophages during the transition period and estimate the temporal range of the phenotypic shift.
Ultimately, this work provides a comprehensive understanding of macrophage dynamic plasticity during wound healing and establishes a quantitative framework for analyzing macrophage behavior \textit{in vivo}.

\section*{Results}

\subsection*{Comparative Analysis of Morpho-Kinetic Behaviors in M1 and M2 Macrophages}

\begin{figure}[!ht]
\centering
\includegraphics[width=\linewidth]{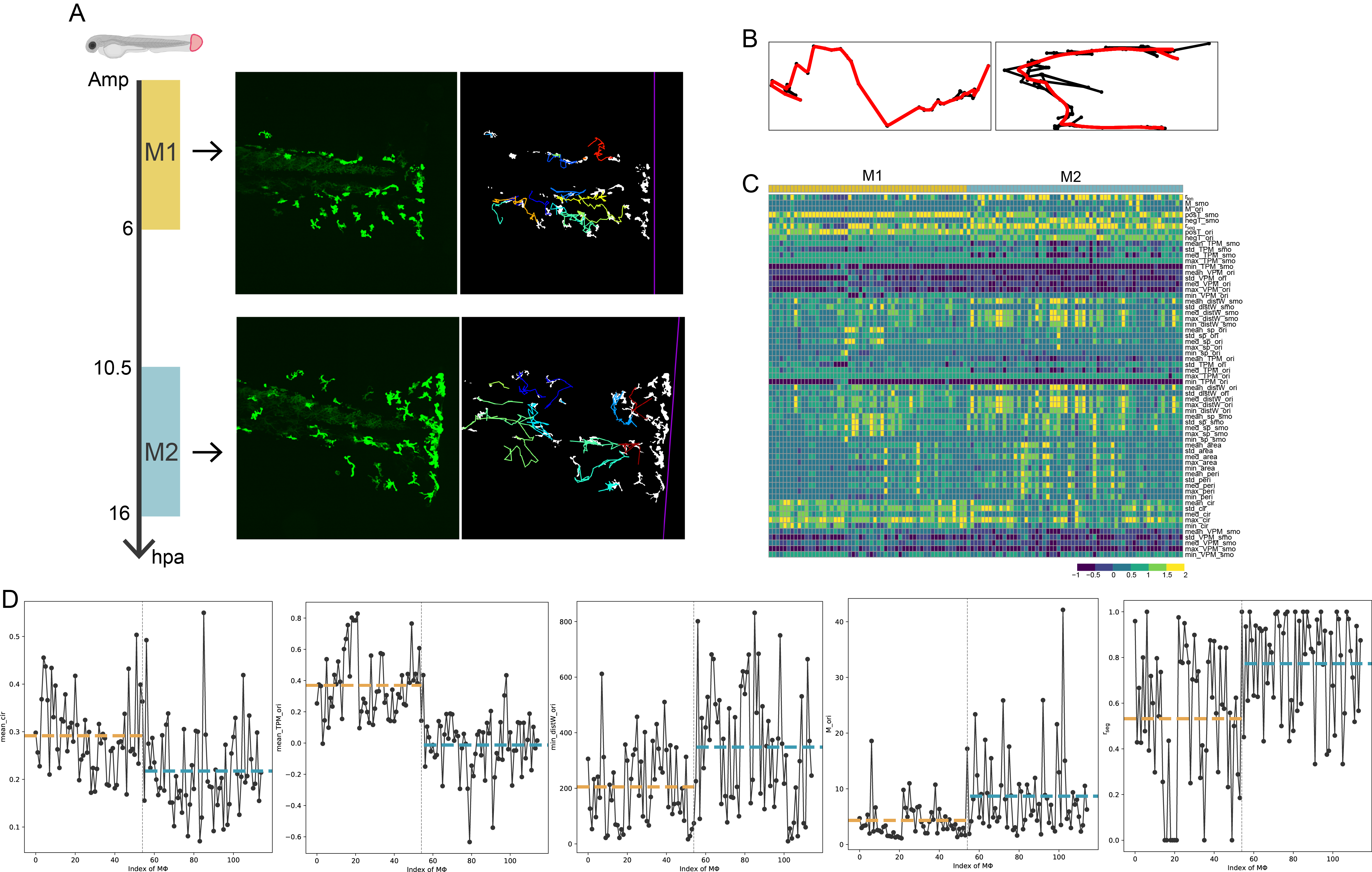}
\caption{\textbf{Quantitative Analysis of Morphology and Movement in M1 vs. M2 Macrophages.} A: Examples of macrophage segmentation and tracking from annotated M1 and M2 videos, with vertical violet lines indicating the wound site. B: Smoothed trajectories (red) overlaid on original extracted trajectories (black). C: Feature matrix of 63 morpho-kinetic features (rows) across trajectories (columns). For visualization purposes, feature values were scaled to a range of 0-2 for positive-only features and –1 to 1 for features with both positive and negative values. M1 and M2 video labels are indicated above the matrix. D: Comparison of features between M1 and M2 macrophages (left to right): mean circularity (mean\_cir), mean Tangent Projection Metric (mean\_TPM\_ori), minimum distance to the wound (min\_distW\_ori), meander ratio (M\_TPM\_ori), and proportion of random-like motion ($\text{r}_\text{seg}$). Abbreviations for the feature names are described in detail in the Supplementary Information. The horizontal axis represents macrophage trajectory indices, with M1 on the left and M2 on the right of the vertical divider. Dashed yellow and blue lines indicate the average values for M1 and M2, respectively. Features with ``ori'' suffix were computed from original (unsmoothed) trajectories.}
\label{fig:M1M2}
\end{figure}

Using annotated microscopy videos of M1 (up to 6 hours post-amputation, hpa) and M2 (10.5--16 hpa) macrophages, we compared their morpho-kinetic features (see Materials and Methods for details on video annotation and feature extraction). Based on macrophage segmentation, tracking (Figure \ref{fig:M1M2}A), and trajectory smoothing (Figure \ref{fig:M1M2}B), we computed 63 features (Figure \ref{fig:M1M2}C) from 3308 individual macrophages across 55 trajectories in M1 videos and 3034 macrophages across 60 trajectories in M2 videos. Across multiple features, M1 and M2 macrophages displayed distinct characteristics, including differences in cell morphology, movement directionality, proximity to the wound site, trajectory shapes, and patterns of random-like motion.

%\subsubsection*{Morphological Differences}
Mean circularity per trajectory revealed that M1 macrophages generally exhibited higher circularity than M2 macrophages (Figure \ref{fig:M1M2}D, leftmost panel), indicating that M2 macrophages tend to have more elongated and complex shapes. This observation is consistent with previously reported findings \cite{sipka2022macrophages}.

%\subsubsection*{Directional Movement Toward the Wound}
To assess movement directionality, we used the ``Tangent Projection Metric'' (TPM); a detailed mathematical definition is available in the Supplementary Information. A TPM value closer to 1 indicates movement toward the wound (along the normal vector of the wound line), while a value closer to -1 indicates movement away. We computed TPM for the original trajectories, indicated by the suffix ``ori'' in the figure. As shown in the second panel of Figure \ref{fig:M1M2}D, the mean TPM for most M1 trajectories was positive, whereas many M2 trajectories had negative TPM values. The average TPM thus showed a clear difference between the two phenotypes, suggesting that M1 macrophages exhibit stronger directional migration toward the wound.

%\subsubsection*{Proximity to the Wound}
We also examined how closely macrophages approached the wound. The mean distance can be misleading—for example, when a macrophage moves from a distant location to the wound or vice versa, the resulting average may fall in the mid-range despite the cell having reached the wound. Therefore, we used the minimum distance in each trajectory as a more reliable indicator of wound proximity. In the third panel of Figure \ref{fig:M1M2}D, the minimum distances from the wound site for the original trajectories (indicated by the suffix ``ori'') are shown for each M1 and M2 trajectory. The average minimum distance (dashed lines) was smaller for M1 macrophages than for M2, suggesting that M1 macrophages are more frequently located near the wound site.
Note that although a macrophage that begins near the wound and migrates away would also yield a low minimum distance, this measure still reflects the overall accumulation tendency near the wound.

%\subsubsection*{Linearity of Motion}
Trajectory linearity was assessed using the meander ratio, defined mathematically in the Supplementary Information. A value of 1 indicates a perfectly straight path, while higher values reflect increasingly winding trajectories. In the fourth panel of Figure \ref{fig:M1M2}D, the original trajectories from M1 showed lower meander ratios than M2, implying that M1 macrophages move in a more directed and linear manner.

%\subsubsection*{Random-like Motion Patterns}
Finally, we quantified the portion of each trajectory exhibiting random-like trajectories using the feature, $\text{r}_\text{seg}$. This measurement quantifies the proportion of time a macrophage spends in random-like motion relative to the total duration of its trajectory. Random-like motion is identified by self-intersections within the trajectory, where the macrophage’s path crosses over itself, which can also indicate looping behavior. Higher values suggest that a larger portion of the trajectory involves looping or non-directional movement. As shown in the rightmost panel of Figure \ref{fig:M1M2}D, M2 macrophages had higher values of $\text{r}_\text{seg}$, indicating that they spent more time in patterns of looping motion. This finding is consistent with the higher meander ratios observed in M2 macrophages.

\subsection*{Macrophage Behavior During the Transition Period}
To investigate macrophage behavior during the transition period (TP), we analyzed videos recorded between 6 and 10.5 hpa. We extracted trajectories from this period and classified them as M1-like or M2-like, based on features previously obtained from annotated M1 and M2 videos. For classification, we applied a nonlinear graph diffusion-based method \cite{mikula2023natural,mikula2023natural2} (see Materials and Methods). To optimize the model parameters, we used the feature matrix from the reference M1 and M2 datasets (Figure \ref{fig:M1M2}C) as the training set for the model. The 113 trajectories extracted from the TP, containing 5743 individual macrophages across all time frames, were then classified by the trained model.
%To investigate the behavior of macrophages during the transition period (TP), we analyzed vidoes, 6-10.5 hpA, and classified the trajectories extracted in this period which trajectories are close to M1 or M2 based on features extracted from M1 and M2 videos in the previous section.
%we applied the nonlinear graph-diffusion-based classification \cite{mikula2023natural} (see Materials and Methods) using the feature matrix obtained from annotated M1 and M2 macrophage data (Figure \ref{fig:M1M2}C). 
%The total number of individual macrophages is 5743 from 113 trajectories in this analysis.

\begin{figure}[!ht]
\centering
\includegraphics[width=\linewidth]{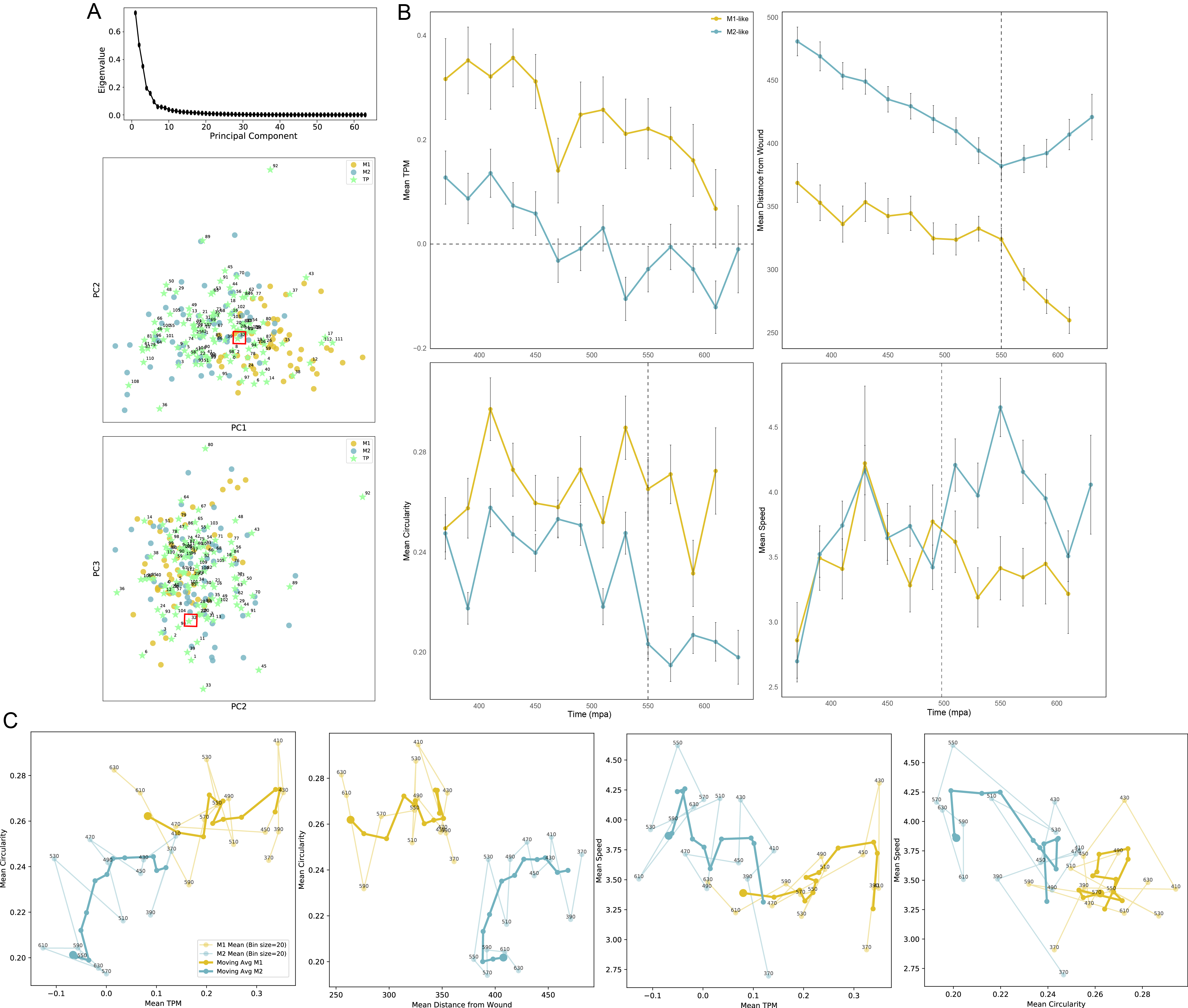}
\caption{\textbf{Analysis of Macrophage States During the Transition Period.} A: PCA-based analysis: eigenvalue decay (top), trajectory projection in PC1-PC2 (middle) and PC2-PC3 (bottom) planes. A TP trajectory better separated in PC2–PC3 is marked with a red square (32$^{\text{nd}}$ point). B: Temporal trends of key features in classified M1-like and M2-like trajectories (20-minute bins): mean TPM, distance to the wound, circularity, and speed. C: Pairwise relationships between features over time, showing mean values for the following pairs: TPM vs. circularity, circularity vs. distance from the wound, TPM vs. speed, and circularity vs. speed. The last temporal point of the moving average is marked slightly larger.}
\label{fig:TP}
\end{figure}

%\subsubsection*{Macrophage Classification in Reduced Feature Space}
The classification was performed in a reduced feature space, where the original 63 features were projected onto 6 dimensions based on the pattern of the eigenvalue decay from principal component analysis (PCA), as shown in the top panel of Figure \ref{fig:TP}A.
The second panel of Figure \ref{fig:TP}A shows that PCA of labeled M1 and M2 videos results in a largely distinct separation along the first two principal components (PC1-PC2), with some overlap. Interestingly, some trajectories in the TP located in the overlapping region of the PC1-PC2 plane (e.g., the 32$^{\text{nd}}$ trajectory, highlighted with a red square) become distinguishable when additional dimensions, such as PC2-PC3, are considered. Incorporating additional dimensions improved classification performance, with a training accuracy of 90.4\%. As a result, the 113 trajectories from the TP were classified as M1-like (37 trajectories), M2-like (75 trajectories), and unclassified (1 trajectory).

%\subsubsection*{Temporal Trends in Morpho-Kinetic Features}

Temporal dynamics of several key features were analyzed across the classified M1- and M2-like trajectories (Figure \ref{fig:TP}B). Trajectory data were binned into 20-minute intervals, and the mean feature values were calculated for each bin, considering only those time bins containing a minimum of 50 individual macrophages. The top-left panel of Figure \ref{fig:TP}B displays the mean TPM, indicating movement directionality relative to the wound. Around 460 minutes post-amputation (mpa), the classified M2 trajectories begin to exhibit TPM $\le$ 0, suggesting reduced migration toward the wound, while classified M1-like trajectories maintain directional movement. The top-right panel shows mean distance to the wound, where the classified M2-like begin to migrate away from the wound around 550 mpa. 
In the bottom-left panel, the circularity sharply declines within the same time bin ($\sim$ 550 mpa), suggesting a morphological shift toward more elongated cell shapes. Additionally, M1-like trajectories exhibit higher speeds prior to 500 mpa (in the bottom-right panel), but after this time, M2-like trajectories exceed them in motility.

%\subsubsection*{Feature Relationships in Classified M1 and M2 During the Transition Period}
To further examine relationships between morpho-kinetic features over time, we generated two-dimensional feature plots for each time bin, represented by semi-transparent points and lines annotated with time bin labels (Figure \ref{fig:TP}C). To enhance the visualization of temporal trends, moving averages were calculated using a three-bin sliding window (i.e., averaged over every three consecutive time bins) and are shown as thicker lines in Figure \ref{fig:TP}C.
The left two panels show that the mean circularity of M2-like trajectories during the TP begins to decline as they reduced their active migration toward the wound.
The right two panels show distinct speed patterns between classified M1-like and M2-like macrophages during the TP: M2-like macrophages accelerate during the phase when active migration to the wound decreases, whereas M1-like macrophages slow down. Notably, the mean speed of classified M2-like macrophages appears to be inversely correlated with circularity, as shown in the rightmost panel.

\subsection*{Non-M1 Macrophages in Comparison with Activated M1 States}
\begin{figure}[!ht]
\centering
\includegraphics[width=\linewidth]{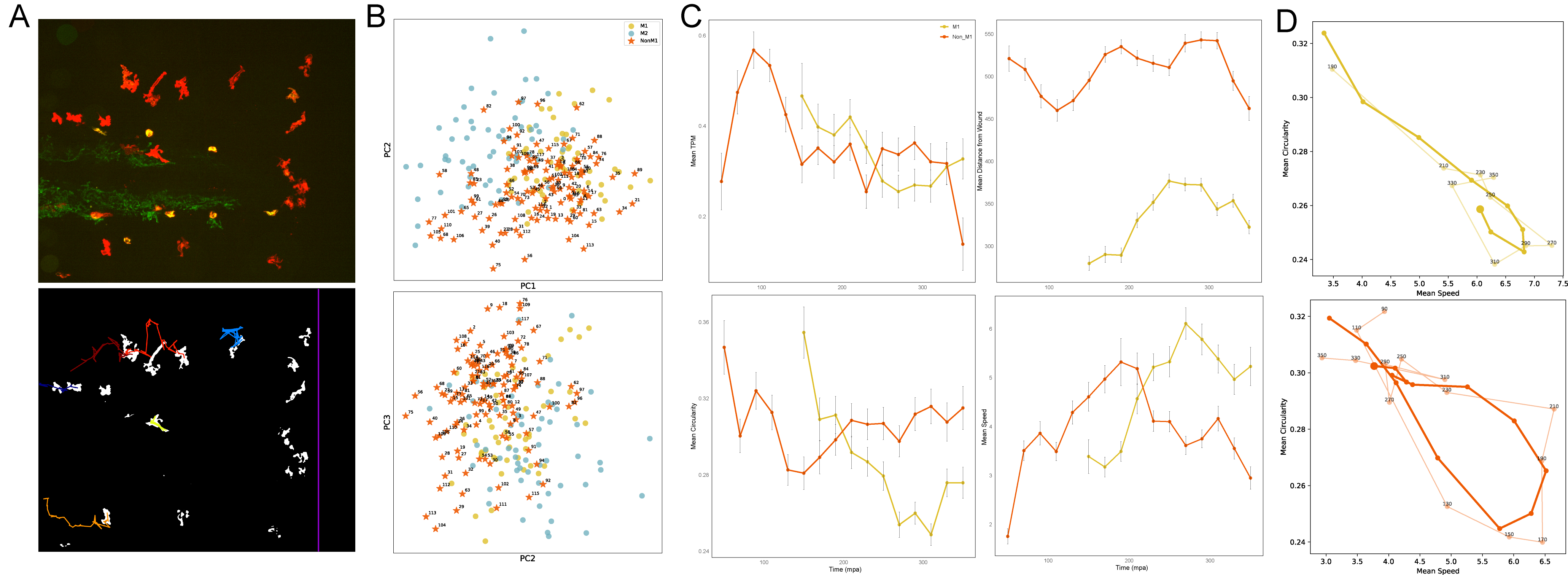}
\caption{\textbf{Morpho-kinetic comparison of Non-M1 and M1 macrophages} A: Example images showing segmentation and trajectories of Non-M1 macrophages, with the vertical violet line indicating the wound site. B: Projection of Non-M1 trajectories onto the PCA planes derived from labeled M1 and M2 data. C: Temporal trends of selected features for M1 and Non-M1 macrophages: mean TPM (top-left), mean distance to the wound (top-right), mean circularity (bottom-left), and mean speed (bottom-right). D: Pairwise feature relationship over time: mean speed versus circularity for M1 (top) and Non-M1 (bottom). The last temporal point of the moving average is marked slightly larger.}
\label{fig:M1NonM1}
\end{figure}

The morphologies and trajectories of non-M1 macrophages, defined as macrophages that do not express the tnfa:GFP-f (green) transgene, were extracted as shown in Figure \ref{fig:M1NonM1}A (see Materials and Methods and Supplementary Figure 1 for details of the extraction process). We extracted 5266 individual macrophages from 118 trajectories.
A total of 63 morpho-kinetic features were computed and projected into PCA planes derived from labeled M1 and M2 macrophage data as shown in Figure \ref{fig:M1NonM1}B. In this PCA plane, many Non-M1 trajectories are positioned closer to the M1 region, particularly along the PC1-PC2 plane, suggesting that Non-M1 macrophages generally exhibit morpho-kinetic characteristics more similar to M1 than to M2. However, several key features distinguish Non-M1 from activated M1 macrophages.
As shown in Figure \ref{fig:M1NonM1}C, both M1 and Non-M1 macrophages migrate toward the wound (similar TPM range); however, Non-M1 macrophages tend to remain at significantly greater distances from the wound site. In terms of morphology, Non-M1 macrophages maintain relatively stable circularity over time, whereas M1 macrophages become progressively more elongated (lower circularity) as they migrate. Additionally, M1 macrophages show a gradual increase in speed over time, whereas Non-M1 macrophages accelerate shortly after amputation but slow down later.
To explore relationships between features, Figure \ref{fig:M1NonM1}D shows the two-dimensional plot of mean speed versus circularity for M1 and Non-M1 macrophages. In general, both modes show an inverse relationship between circularity and speed. In Non-M1 macrophages, circularity tends to decrease as speed increases during the early phase following amputation; however, circularity begins to rise around 170 mpa, accompanied by a decrease in speed.  In contrast, M1 macrophages exhibit progressively higher speeds and decreasing circularity over the observed period.

\subsection*{Macrophage Behavior in the Absence of Wound Signal}

\begin{figure}[!ht]
\centering
\includegraphics[width=\linewidth]{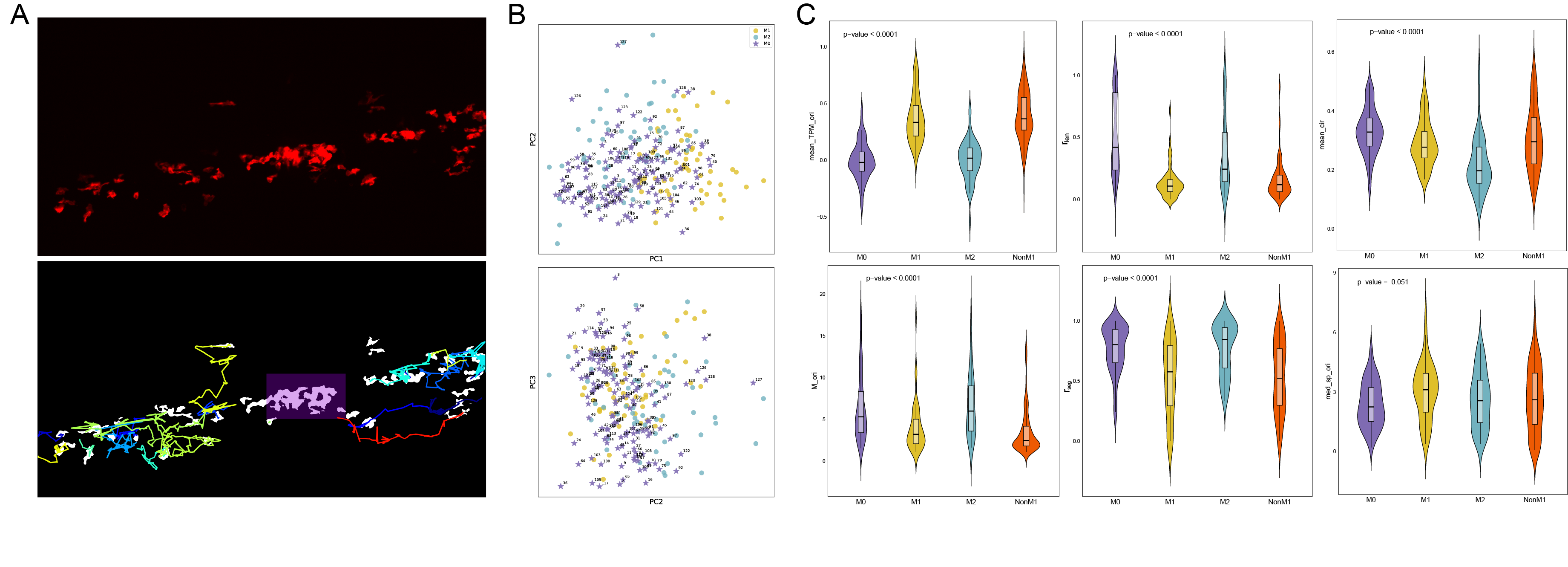}
\caption{\textbf{Morpho-kinetic analysis of M0 macrophages in the absence of wound signal.} A: Example images showing segmentation and trajectories of M0 macrophages. Transparent violet regions (bottom) indicate manually defined macrophage accumulation zones. B: Projection of M0 trajectories onto the PCA planes derived from labeled M1 and M2 macrophages. C: Comparison of M0 and other macrophage modes across selected features: mean TPM (top-left), mean $\text{r}_\text{len}$ (top-middle), mean circularity (top-right), meander ratio (bottom-left), mean $\text{r}_\text{seg}$ (bottom-middle), and median speed (bottom-right). Abbreviations for the feature names are described in detail in the Supplementary Information. Features with ``ori'' suffix were computed from original (unsmoothed) trajectories.}
\label{fig:M0}
\end{figure}

We analyzed the behavior of M0 macrophages, defined as macrophages not exposed to a wound signal, from a morpho-kinetic perspective. The steps of video processing were consistent with previous analyses; however, only the mfap4:mCherryF (red) channel was used (top panel of Figure \ref{fig:M0}A), since no inflammation was induced therefore the tnfa:GFP-f (green) transgene was not expressed. %(data not shown or See supplementary... !!this will be decided later from Mai!!) 
In these unwounded conditions, macrophages tended to accumulate in specific regions (see Supplementary Figure 2). For reasonable analysis, we excluded trajectories in which cells remained within such an accumulation zone for more than five consecutive time frames (see bottom panel of Figure \ref{fig:M0}A and Materials and Methods for details). As a result, 9917 individual M0 macrophages were extracted with 132 trajectories.
To compute features related to the wound site, such as movement direction (TPM) and distance from the wound, a reference line corresponding to the site of wound is required. However, since M0 macrophages are from unwounded cases, we defined a vertical line closely matching the location of the caudal fin fold to maintain consistency with the wounded condition, as amputations in this study were performed in the caudal fin fold.
Using 63 morpho-kinetic features, we projected the M0 trajectories onto the PCA planes derived from labeled M1 and M2 macrophages. As shown in Figure \ref{fig:M0}B, many M0 trajectories cluster near the M2 region in the PC1-PC2 plane, suggesting that M0 macrophages share greater similarity with M2 than with M1 in terms of overall morpho-kinetic behavior.

In Figure \ref{fig:M0}C, the violin plots show the similarities and differences between M0 macrophages and other activation states, and also provide a comprehensive overview of the distributional differences across all activation states.
In the top-left panel, TPM (computed from original trajectories) for M0 is close to zero, and many M0 trajectories are distributed near this value, indicating a lack of clear directionality. The median value of mean TPM for M0 closely resembles that of M2 macrophages. However, the M2 macrophages exhibits a longer tail in the negative TPM range, reflecting the active reverse migration described earlier (see first Results section).
In the bottom-left panel, the meander ratio was computed from original trajectories. For M0 macrophages, it is higher than for M1 and Non-M1 and more closely resembles that of M2, indicating more curved and less directed trajectories.
The middle panels (top and bottom) show results for two features capturing random-like movement: $\text{r}_\text{len}$, representing the proportion of a trajectory’s length involved in self-intersecting trajectories, and $\text{r}_\text{seg}$, representing the proportion of time spent in such patterns. Both measurement are significantly higher in M0 macrophages than in M1 and Non-M1, and are similar to M2 values, further indicating a predominance of random patterns of circuitous motion.
Despite these similarities with M2 macrophages, M0 macrophages also exhibit distinct features, as shown in the rightmost panels. In the top-right panel, M0 macrophages show the highest mean circularity among all modes, indicating the most rounded morphology, in contrast to the highly elongated shapes observed in M2 macrophages. Furthermore, in the bottom-right panel, M0 macrophages exhibit the lowest median of speed (computed from original trajectories) among all activation states, suggesting a relatively inactive or weakly migratory state.

\section*{Discussion}
In this study, we investigated distinct behavioral modes of macrophages during wound healing in a zebrafish model. Using a comprehensive analysis pipeline, we computed 63 morpho-kinetic features to characterize macrophage behavior in a high-dimensional feature space (see Supplementary Information, Supplementary Figure 6, for the full analytic workflow).

We quantitatively compared M1 and M2 macrophages across several aspects, including morphology, movement directionality, proximity to the wound, and trajectory shape. Our results showed that M2 macrophages are more elongated, exhibit less directed migration toward the wound (and in some cases reverse migration), maintain higher meander ratios (indicating circuitous paths), and tend to remain farther from the wound site. These findings suggest that M2 macrophages may be responding to a weaker or diminishing chemoattractant signal, resulting in less targeted movement toward the wound.

Using the features extracted from annotated M1 and M2 data, we classified macrophage trajectories from the transitional period (6--10.5 hours post-amputation, hpa) into M1-like or M2-like groups.
The classification indicates that the M1-like and M2-like groups closely resembled the annotated M1 and M2 trajectories, respectively, in terms of morpho-kinetic properties averaged over the observation period for each trajectory. The extracted trajectories during the transition period (TP) originate from macrophages that have been or are currently expressing the tnfa:GFP-f transgene. The classifier was trained to distinguish only between M1-like and M2-like behaviors based on these average features within the TP; therefore, the assigned labels reflect the dominant phenotype of each trajectory rather than the timing of the M1-to-M2 switch. As a result, some trajectories classified as M1-like may exhibit M2-like behavior in the later part of the TP, while some M2-like trajectories may include M1-like behavior in the early part of the TP. To better capture the temporal dynamics of the M1-to-M2 transition, we analyzed the classified groups by dividing their trajectories into time bins. This allowed us to investigate temporal patterns of morpho-kinetic behavior within each group, providing a more detailed view of their dynamic changes over time. With this temporal analysis of the classified M1-like and M2-like groups, we were able to estimate the approximate timing of the M1-to-M2 shift. Classified M2-like macrophages begin to show TPM values $\le$ 0 (indicating loss of directed movement toward the wound) around 460 minutes post-amputation (mpa), with reverse migration observed after ~550 mpa. Therefore, the time range of 460–550 mpa likely corresponds to the point at which M1 macrophages begin transitioning to the M2 mode. Moreover, pairwise feature analysis revealed that M2-like macrophages exhibit increased speed and elongation during reverse migration, providing further insight into the dynamics of this shift.

We also analyzed Non-M1 macrophages that do not express TNF. Although most Non-M1 macrophages clustered near M1 macrophages in the PCA planes based on morpho-kinetic features, they exhibited a significantly greater distance from the wound, indicating that they were located farther away. This suggests that TNF transcriptional activation, indicated by tnfa:GFP-f expression, is spatially associated with wound proximity. Finally, we examined M0 macrophages (those not exposed to wound signals). These macrophages displayed no clear directionality, exhibited similar patterns of random-like movement to M2 macrophages, but maintained high circularity and low speed, indicating a rounded and minimally migratory phenotype.

Together, our findings provided a landscape of macrophage behavior during zebrafish wound healing based on \textit{in vivo} imaging. The 63 morpho-kinetic features captured distinct behavioral characteristics across different macrophage modes. Several \textit{in vitro} studies have also explored macrophage classification using morphology and movement. For example, in a study using primary murine bone marrow-derived macrophages \cite{kesapragada2024deep}, cells were maintained in the M0 state, or polarized toward M1 or M2 states, and they were tracked under time-lapse imaging. A deep learning model trained only on cell trajectories achieved over 90\% classification accuracy, demonstrating that kinetic patterns are informative for distinguishing macrophage modes.  Also, the authors interpreted characteristic migratory behaviors for each macrophage subtype e.g., M0 macrophages were described as ``jiggling in place'', which is consistent with our observation of low-motility M0 macrophages \textit{in vivo}. Another study focused on morphological features, using macrophages differentiated from human monocytes \cite{rostam2017image}. The authors extracted shape descriptors (e.g., area, eccentricity, etc.) and were able to classify M1 and M2 cells, and distinguish them from naive macrophages and monocytes, with approximately 90\% accuracy. This indicates that cell morphology differs significantly between polarization states and can be significant features to distinguish them. In addition, a study using two-photon fluorescence lifetime imaging (2P-FLIM) \cite{neto2022non} demonstrated that M1 and M2 macrophages exhibit distinct metabolic dynamics. By analyzing fluorescence lifetime parameters reflecting metabolic activity, and applying a random forest classifier, with achieved score up to 0.94. This work highlights that metabolic behavior, similarly as motility and morphology, is a discriminative feature of macrophage polarization. These studies collectively show that macrophage polarization is different not only in gene expression, but also in morphological, kinetic, and metabolic characteristics, implying each of which can be used as a key descriptor for classification. Our morpho-kinetic analysis, uniquely performed \textit{in vivo}, aligns with this view and demonstrates that shape and migratory behavior can be effective indicators of macrophage modes.

As future work, more comprehensive, multi-dimensional analyses that integrate multiple features simultaneously could provide a richer understanding of macrophage behavior across modes. In our current study, we observed clear behavioral differences across modes, but our interpretations largely focused on individual or pairwise features. Additionally, the estimated timing of phenotypic switching (from M1 to M2) reflects an overall trend rather than a precise transition point, due to classifying trajectories over their full duration rather than specific time segments. Therefore, developing analytic methods capable of identifying the precise timing of phenotypic shifts at finer temporal resolution will be a valuable direction to extend this study.

%For labelling the trajectories for each mode, the authors used the geometrical properties where M0 cells are circular, M1 cells are protruded, M2 cells are elongated.

%However, our interpretation largely focused on individual or pairwise feature comparisons. Additionally, the estimated timing of phenotypic switching (from M1 to M2) reflects an overall trend rather than a precise transition point, due to classifying trajectories over their full duration rather than specific time windows. 

\section*{Materials and Methods}

\subsection*{Ethics statement}
Fish husbandry, embryo collection and animal experiments were conducted by well-trained and authorized staff at LPHI laboratory (CNRS/UMR5294), University of Montpellier, according to European Union guidelines for handling of laboratory animals (\url{https://ec.europa.eu/environment/chemicals/lab_animals/index_en.htm}) and were approved by the Comité d’Ethique pour l’Expérimentation Animale under reference CEEA-LR- B4-172-37. Developmental stages used were between 3 days post-fertilization (dpf) and 4 dpf. Specimen were euthanized using an anesthetic overdose of buffered tricaine.

\subsection*{Data Acquisition and Macrophage Annotation}
The datasets used in this study consist of \textit{in vivo} videos of migrating macrophages in the tail region of transgenic Tg(mfap4:mCherry-F)ump6 ; Tg(tnfa:GFP-F)ump5 zebrafish larvae, at 3 dpf. To induce a wound response, larvae were first anaesthetized in zebrafish medium supplemented with $200$ $\mu g/ml$ tricaine (ethyl 3-aminobenzoate methanesulfonate, MS-222 sigma \# A5040). Then, the caudal fin fold were amputated with a sterile scalpel, posterior to muscle and notochord and larvae were either transferred in their medium at 28$\,^\circ\mathrm{C}$ until later imaging or mounted in 1\% low-melting point agarose approximately 20 minutes after amputation (to allow partial wound closure), previously described \cite{sipka2022macrophages}. Intra-vital imaging was performed using on ANDOR CSU-W1 confocal spinning disk on an inverted NIKON microscope (Ti Eclipse) with ANDOR Neo sCMOS camera (20x air/NA 0.75 objective); lasers 488 $nm$ (GFP) and 561 $nm$ (mCherry). Image stacks for time-lapse movies were acquired at 28$\,^\circ\mathrm{C}$ with following setups: Time steps and time ranges are indicated in Supplementary Table 1, z step was 1 to 3 $\mu m$. Macrophages were identified by the expression of the mfap4:mCherry-F transgene, allowing the production of a membrane-localized red fluorescent protein indicating cell identity. M1 activation of macrophages was identified by the expression of the tnfa:GFP-f transgene, allowing the production of a membrane-localized green fluorescent protein in cell producing TNF, a known marker of M1. For the analysis, 4D files generated from time-lapse acquisitions were projected into two dimensions using the maximum intensity projection. Macrophage behavior observed in the videos was annotated into four distinct modes, as detailed in Table 1 of the Supplementary Information. As no specific marker for M2 macrophages was available in this experimental system, M1 and M2 annotations were assigned based on the time post-amputation, reflecting the time-dependent polarization from M1 to M2 states during wound healing in zebrafish. The annotation of macrophages considered both activation of the tnfa:GFP-f transgene and the time after amputation. M1 expressed tnfa:GFP-f and were present during the first 6 hours post-activation (hpa), corresponding to the pro-inflammatory phase with macrophages up-regulating cytokines, chemokines and key pro-inflammatory pathways, meanwhile increasing levels of genes of glycolysis \cite{BegonPescia2025}. M2 were defined as macrophages transitioning from M1, observed between 10.5 and 16 hpa, a time window corresponding to inflammation resolution; macrophages from this period down-regulated pro-inflammatory pathway while expressing metalloproteinases \cite{BegonPescia2025}. Non-M1 were macrophages that did not show TNF expression within the first 6 hpa; and M0 were macrophages in homeostasis, not stimulated by a wound. Additionally, macrophages showing TNF expressing during the transition period (6–10.5 hpa) were labeled as TP.

\subsection*{Macrophage Segmentation}
Macrophage segmentation was performed using a multi-step workflow consisting of space-time filtering, local Otsu thresholding, and the Subjective Surface Segmentation (SUBSURF) method, as described in Park et al. \cite{park2023segmentation}.
\subsubsection*{Space-Time Filtering}
Space-time filtering was applied to reduce noise while preserving macrophage signals by considering their temporal coherence. For a macrophage video with time slices on the interval [$0,\theta_{F}$], where $\theta$ represents a particular time slice, the space-time filtering is governed by
\begin{equation} \label{sp_filter}
\frac{\partial u}{\partial t} = clt(u)\nabla \cdot \bigl(g(|\nabla G_{\sigma}*u|)\nabla u\bigr).
\end{equation}
Here, $t$ represents the scale (amount of filtering) and $u(t,x_1,x_2,\theta)$ is the unknown real function defined on $[0,T_{F}] \times \Omega \times [0,\theta_{F}]$, with $\textbf{x}=(x_1,x_2) \in \Omega \subset R^{2}$. 
The $clt(u)$ function is defined as \cite{sarti1999nonlinear}
\begin{equation} \label{clt}
clt(u)=\min_{\boldsymbol{w_1}, \boldsymbol{w_2}}\frac{1}{(\Delta \theta)^2} \bigl(|<\nabla u,\boldsymbol{w_1}-\boldsymbol{w_2}>|+|u(\mathbf{x}-\boldsymbol{w_1}, \theta-\Delta \theta)-u(\mathbf{x},\theta)|+|u(\mathbf{x}+\boldsymbol{w_2}, \theta+\Delta \theta)-u(\mathbf{x},\theta)|\bigr),
\end{equation}
where $\boldsymbol w_1$ and $\boldsymbol w_2$ are arbitrary vectors in 2D space, $\Delta\theta$ is the time increment between discrete time slices, and $<\boldsymbol{a},\boldsymbol{b}>$ denotes the Euclidean scalar product. 
The edge detector function $g$ is defined as
\begin{equation} \label{g_function}
g(s)=\frac{1}{1+Ks^2}, \ K>0,
\end{equation}
where $K$ controls the sensitivity to $s$. $G_\sigma$ is a Gaussian function with variance $\sigma$ used for pre-smoothing by convolution.
\subsubsection*{Local Otsu Thresholding}
After filtering, we applied the local Otsu's method to identify approximate macrophage regions by adaptively calculating thresholds for each pixel. 
The resulting binarized images were obtained based on
\begin{equation} \label{binarized}
  B(i,j)=
	\begin{cases}
	1,  I(i,j) > T^{*}_{i,j} \ \text{and Equation \ref{local_otsu_eval}} \enspace \text{is fulfilled}  \\
	0, \text{otherwise,} \\ 
 \end{cases}  
\end{equation}
\begin{equation} \label{local_otsu_eval}
    \frac{|\mu_{0}(T^{*}_{i,j})-\mu_{1}(T^{*}_{i,j})|}{\mu_{0}(T^{*}_{i,j})} > \delta,
\end{equation}
where $I(i,j)$ is the image intensity of a pixel $(i,j)$.
The optimal threshold $T^{*}_{i,j}$ maximizes the between-class variance

\begin{equation} \label{localotsu_variance}
    \begin{aligned}
    \sigma_{B}^{2}(T^{*}_{i,j})&=\max_{0 \leq T_{i,j} < L} \sigma_{B}^{2}(T_{i,j}), \\
    \sigma_{B}^{2}(T_{i,j})&=\frac{\bigl(\mu_{\text{tot}}\omega_{0}(T_{i,j})-\mu_{0}(T_{i,j})\omega_{0}(T_{i,j})\bigr)^{2}}{\omega_{0}(T_{i,j})\bigl(1-\omega_{0}(T_{i,j})\bigr)},
    \end{aligned}
\end{equation}

with the following terms defined as
\begin{equation} \label{localotsu_w}
\begin{aligned}
\omega_{0}(T_{i,j})&=\sum^{T_{i,j}}_{r=0}p_{r}, \quad
\mu_{0}(T_{i,j})&=\frac{1}{\omega_{0}(T_{i,j})}\sum^{T_{i,j}}_{r=0}rp_{r}, \quad
\mu_{\text{tot}}&=\sum^{L}_{r=0}rp_{r},
\end{aligned}
 \end{equation}
where $p_{r}$ is the probability of intensity $r$ in the local gray-level histogram, calculated as $p_{r}=n_{r}/N$, with $n_{r}$ being the number of pixels with intensity $r$ and $N$ the total number of pixels in the local window. 
\subsubsection*{Subjective Surface Segmentation (SUBSURF)}
To refine macrophage boundaries and remove residual noise, the SUBSURF method \cite{sarti2000subjective} was applied to each 2D frame independently.
The SUBSURF method is described by
\begin{equation} \label{subsurf}
\frac{\partial u}{\partial t}=|\mathbf{\nabla}u|\mathbf{\nabla}\cdot\biggl(g\frac{\mathbf{\nabla}u}{|\mathbf{\nabla}u|}\biggr),
 \end{equation}
where $u$ is a level set function, $g = g\left( \left| \nabla G_\sigma * I_0 \right| \right)$ as previously defined, and $I_0$ denotes the original image. The SUBSURF equation was solved for $u(t,x)$ with $(t,x) \in [0, T_S] \times \Omega$, where $\Omega \subset \mathbb{R}^2$.
\subsubsection*{Parameter Settings and Post-processing}
All segmentation parameters followed those in the original article \cite{park2023segmentation}, except for $\delta$ in the local Otsu thresholding. Specifically, the parameter $\delta$ was set to $0.4$ for M1 and TP videos, $0.3$ for M0 videos, and $0.5$ for M2 videos.
As a post-processing step, we applied a weighted dilation and erosion approach \cite{park2024automated} to complete fragmented macrophage segments, described by
\begin{equation} \label{levelset_eq}
    \frac{\partial u}{\partial t} \pm v_{i,j} |\nabla u| = 0,
\end{equation}
where $v_{i,j} = \frac{1}{2}(\Tilde{S}_{i,j} + I_{i,j})$, with $\pm$ denoting erosion ($+$) and dilation ($-$), and $\Tilde{S}_{i,j} = 1 - |I_{i,j} - T_{i,j}^{*}|$. This process involved 10 iterations each of dilation and erosion across all videos.

\subsection*{Segmentation of Non-M1 Macrophages}
Identification of Non-M1 macrophages, those present in mfap4:mCherryF (red channel) but lacking tnfa:GFP-f (green channel) activation, were identified by segmenting both green and red fluorescence channels using the same procedure, Equations \eqref{sp_filter}--\eqref{levelset_eq}. The red channel used a threshold parameter $\delta = 0.3$, with all other parameters matching those of the green channel (M1 videos).
Non-M1 macrophages were then detected by assessing the overlap between segmented regions in the red and green channels across all time frames. Segmented macrophages that were present exclusively in the red channel, without any corresponding overlap in the green channel, were annotated as Non-M1.
A detailed overview of this process is illustrated in Supplementary Figure 1 of the Supplementary Information.

\subsection*{Macrophage Tracking}
\subsubsection*{Approximated Centers and Trajectory  Extraction}
For tracking, we followed the approach outlined in Park et al. \cite{park2023segmentation}. The process began by computing the approximated center of each macrophage through the solution of the Eikonal equation from the boundary of segmented regions.
\begin{equation}
\begin{aligned} 
|\nabla d(\mathbf{x})| &= 1, \quad & \mathbf{x} &\in \Omega, \\
d(\mathbf{x}) &= 0, \quad & \mathbf{x} &\in \partial \Omega,
\end{aligned}
     \label{eq:dist}
\end{equation}
where $\Omega$ denotes a segmented macrophage, $\partial \Omega$ is its boundary, and $| \cdot |$ is the Euclidean norm in $\mathbb{R}^2$. The center was defined as the pixel with the maximum distance to the boundary (i.e., the most interior point).
Next, partial trajectories were extracted from temporally overlapping segmented macrophages. These trajectories were then connected based on the direction of movement.
To determine whether two partial trajectories should be connected, we estimated the cell center position $\mathbf{r}_{\text{es}}$ just before or after the endpoint of a trajectory by extrapolating its path using the local slope, representing the direction of motion. A connection was established if the estimated point was sufficiently close to the endpoint $\mathbf{r}_{\text{e}}$ of another trajectory, satisfying the condition:
\begin{equation}
\begin{aligned}
|\textbf{r}_{\text{es}}-\textbf{r}_{\text{e}}| \leq \Delta r.
\end{aligned}
     \label{cond_connect}
\end{equation}
Additionally, connections were evaluated based on whether the estimated center $\textbf{r}_{\text{es}}$ was located near the start or end point of another trajectory across multiple consecutive time frames. Two trajectories were connected if both of the following conditions were satisfied.
\begin{equation}
\begin{aligned} 
|\textbf{r}_{\text{es}}-\textbf{r}_{\text{j}}| \leq \Delta r_{2} \ \text{and} \enspace \Theta_{c} \leq \Delta r_{\theta},
\end{aligned}
     \label{cond_connect2}
\end{equation}
where $\textbf{r}_{\text{j}}$ represents another trajectory close to the estimated point $\textbf{r}_{\text{es}}$, and $\Theta_{c}$ denotes the number of common time frames between trajectories considered for connection. For all videos analyzed in this study, we set $\Delta r_{\theta}=5$.
The spatial connection parameters $\Delta r$ and $\Delta r_2$ were adjusted depending on the video type.
For M1 macrophages, we set $\Delta r=60$ and $\Delta r_2=120$. For M2 macrophages, Non-M1 macrophages, and M0 macrophages, we used $\Delta r=50$ and $\Delta r_2=100$. For TP videos, we employed intermediate values with $\Delta r=55$ and $\Delta r_2=110$. All remaining parameters were maintained as reported in the original method \cite{park2023segmentation}.
To ensure robustness in tracking, only segmented regions with a maximum distance from the boundary greater than $6.5$ were considered, thereby excluding minor segmented regions.

\subsubsection*{Exclusion of Short and Unreliable Trajectories}
To ensure reliable analysis, only trajectories with a sufficient number of time points, corresponding to at least one hour of continuous movement, were considered. For example, in videos with a temporal resolution of 2.5 minutes per frame, only trajectories containing more than 24 points were included. 
Additionally, due to macrophage accumulation at wound sites, the 2D projection of the data can result in the visual merging of adjacent macrophages. This often leads to segmentation artifacts, where multiple cells appear as a single connected region, potentially leading to inaccurate trajectory interpretation. To mitigate this issue, we manually defined a macrophage accumulation zone (MAZ), indicated by violet rectangles in Figure 2 of the Supplementary Information. Trajectory segments in which macrophages remained within an MAZ for more than five consecutive time frames were excluded for further analysis.

\subsection*{Trajectory Smoothing and Randomness Extraction}
To separately analyze directional and random-like macrophage motion, we applied the trajectory smoothing model proposed by Lupi et al. \cite{lupi2025mathematical}, extracting segments of the random-like motion.
This method is based on evolving curves, formulated through the following partial differential equation.

\begin{equation}
\frac{\partial \mathbf{x}}{\partial t} = -\delta(\mathbf{x}, t)k \mathbf{N} + \lambda(\mathbf{x}, t)\left[(\mathbf{x}_0 - \mathbf{x}) \cdot \mathbf{N}\right] \mathbf{N} + \alpha \mathbf{T}.
\end{equation}
This equation describes the evolution of points $\mathbf{x}$ along the curve through three key components: a curvature-based smoothing term, $-k \mathbf{N}$ that depends on the local curvature $k$; an attracting term, $[(\mathbf{x}_0 - \mathbf{x}) \cdot \mathbf{N}] \mathbf{N}$ that preserves the shape of the original trajectory; and a tangential component, $\alpha \mathbf{T}$ for numerical stability, where $\mathbf{T} $ and  $\mathbf{N}$ denote the unit tangent and normal vectors, respectively.
In this model, the parameters  $\delta(\mathbf{x}, t)$ and $\lambda(\mathbf{x}, t)$ were adaptively selected based on the presence of self-intersections along the curve, and thus depend on both spatial and temporal coordinates.
Trajectory segments that lie within self-intersecting regions were specifically targeted for smoothing.
These segments, referred to as random segments, are defined as portions of the trajectory between two self-intersection points, including all intermediate points. 
%Only random segments containing at least five points are included in the smoothing process.

% If the starting point of a self-intersection is shared by two segments (i.e., it shares the endpoint of both), the segment in the direction of the curve's parametrization is considered the initial segment. Conversely, if the ending point is shared, the segment in the direction opposite to the parametrization is considered the final segment. Only random segments containing at least five points are included in the smoothing process.

\subsection*{Extraction of Morpho-Kinetic Features}
After the aforementioned processing steps, we computed both morphological and kinetic features for individual macrophages from extracted trajectories.

\textbf{Morphological features} were computed from the segmented regions (i.e., connected regions) containing each trajectory point and were analyzed along the full trajectories over time.
For simplicity, we refer to these regions as segmented macrophages; they were derived from binarized images obtained through the segmentation workflow.
For each segmented macrophage, we computed the area, perimeter, and circularity across all time frames. Circularity quantifies the roundness of a shape, where a value of 1 corresponds to a perfect circle. Lower circularity values indicate more elongated or irregular morphologies, capturing greater structural complexity.

\textbf{Kinetic features} focused on four main aspects: trajectory shape, directionality toward the wound, distance to the wound, and speed. Trajectory shape was characterized using both extracted random segments and the meander ratio. Directionality was evaluated by projecting tangent and velocity vectors at each time frame onto the normal vector of the wound site. Distance to the wound was defined as the shortest distance from the macrophage’s current position to the wound boundary. Speed was calculated across all time frames for each trajectory.

The detailed mathematical formulations are described in the Supplementary Information, along with an overview of these features (Table 2).

\subsection*{Classification of Macrophage Trajectories}
To analyze macrophage behavior during the transition period, we classified M1- and M2-like trajectories based on extracted morpho-kinetic features. For this, we used \textbf{NatNet} \cite{mikula2023natural}, a nonlinear graph-diffusion classifier that operates in a reduced feature space.
The model was trained using annotated data consisting of trajectories labeled as M1 or M2, with morpho-kinetic features computed for each trajectory. Optimal model parameters were determined during the training process. After identifying the optimal parameters, the classifier was applied to trajectories from the transition period to assign M1-like or M2-like labels based on their similarity to the annotated trajectories.
The classification model is trained from the annotated data (i.e., features from each extracted trajectories) to find optimal model parameters. Once the optimal parameters are found, the trajectories which we want to classify are applied to the trained model with the optimal parameters.  
Classification is performed by solving the graph diffusion equation, as described below.

\begin{equation} \label{Eq:classification_laplace}
\frac{\partial X(v, t)}{\partial t} = \nabla \cdot \left( g \nabla X(v, t) \right), \quad v \in V(G), \quad t \in [0, T]
\end{equation}

In this equation, $X: G \times [0, T] \rightarrow \mathbb{R}^k$ represents the Euclidean coordinates of vertex $v \in V(G)$ at time $t$, where $t$ is an abstract time variable associated with the graph diffusion process, with $X(v,t) = (x_1(v,t), \dots, x_k(v, t))$ and $k$ being the dimensionality of the reduced feature space $\mathbb{R}^k$. The equation models a diffusion process over the directed graph $G$.

The diffusion coefficient $g$ associated with an edge $e_{uv}$ is defined as

\begin{equation} \label{Eq:classification_g}
g(e_{uv}) = \frac{\varepsilon(e_{uv})}{1 + \sum_{i=1}^{k} K_i l_i^2(e_{uv})}, \quad  K_i \geq 0, \quad i = 1, \dots, k,
\end{equation}
where $\varepsilon(e_{uv})$ depends on the type of diffusion. For edges connecting nodes of the same class, $\varepsilon(e_{uv})$ is positive and represents the forward diffusion that attracts the nodes to each other (typically set to 1). For edges between nodes of different classes, $\varepsilon(e_{uv})$ is assigned a small negative value, representing the backward diffusion that repels the nodes.

Here, $K_i$ are weights assigned to each coordinate $l_i(e_{uv})$ of the vector, $l(e_{uv}) = \left( l_1(e_{uv}), \dots, l_k(e_{uv}) \right)^{\top}$ = $X(v, \cdot) - X(u, \cdot)$ = $\left( x_1(v, \cdot) - x_1(u, \cdot), \dots, x_k(v, \cdot) - x_k(u, \cdot) \right)^{\top}$ for $u, v \in V(G)$.

The accurate classification relies on appropriately tuning the parameters of the diffusion coefficient $g$, particularly the weights $K_i$. 
Once optimal parameters are determined, new nodes (trajectories) can be added to the graph and classified by propagating them through the optimized graph diffusion network. For any new trajectory,  $\varepsilon(e_{uv})$ is set to a positive value such that it is attracted to all nodes of the directed graph, and it is then assigned to the class with which it is most closely associated according to the network dynamics \cite{mikula2023natural2}.

To reduce the dimensionality of the original 63-dimensional feature space, we applied principal component analysis (PCA). We selected the first 6 principal components, as the eigenvalues beyond the 6th component were small and nearly uniform (see Figure \ref{fig:TP}A). The final classification, therefore, was conducted in this reduced 6-dimensional PCA space by solving Equation \eqref{Eq:classification_laplace}.

The optimal diffusion coefficient $g$ was determined using trajectories from annotated videos based on the time period after amputation, each representing either an M1 or M2 trajectory. Morpho-kinetic features from the TP videos were extracted, and trajectories in the TP were then classified by NatNet as M1-like or M2-like.
In this study, the optimal parameters $K_i$, for $i=1,...,6$ in Equation \eqref{Eq:classification_g} were obtained as follows: $K_1=4050, K_2=3800, K_3=50, K_4=550, K_5=3300,$ and $K_6=2800$.

\bibliographystyle{unsrt}
\bibliography{reference}

%\section*{Acknowledgements (not compulsory)}
\section*{Acknowledgements}
This work has received funding from the European Union’s Horizon 2020 research and innovation programme under the Marie Sk\l{}odowska-Curie ITN INFLANET: grant agreement No. 955576, as well as from grants APVV-23-0186, VEGA 1/0249/24, and VV-MVP-24-0116. We thank the Aquatic model facility ZEFIX from LPHI (University of Montpellier) and the imaging facility BioCampus Montpellier Ressources Imagerie (MRI), member of the national infrastructure France-BioImaging supported by the French National Research Agency (ANR-10-INSB-04, 'Investments for the future').

\section*{Author contributions statement}
S.A.P. designed the analysis workflow, tested and performed the data analysis, and wrote the manuscript. G.L. contributed the method of trajectory smoothing and participated in discussions on designing the analysis workflow. R.O. acquired and provided the data and contributed to the biological interpretation of the results. M.K. and A.A.O. contributed the classification method. M.N.C. co-led the project, acquired and provided the data, and contributed to the biological interpretation. K.M. led the project, designed the analysis workflow, and co-wrote the manuscript.
All authors reviewed the manuscript and provided feedback.

\section*{Data availability}
The raw imaging and processed datasets containing extracted macrophage trajectories and morphological features are available at Zenodo: \url{10.5281/zenodo.16319732}.

\section*{Competing interests}
The authors declare no conflict of interest.

%To include, in this order: \textbf{Accession codes} (where applicable); \textbf{Competing interests} (mandatory statement). 

%The corresponding author is responsible for submitting a \href{http://www.nature.com/srep/policies/index.html#competing}{competing interests statement} on behalf of all authors of the paper. This statement must be included in the submitted article file.

% \begin{figure}[ht]
% \centering
% \includegraphics[width=\linewidth]{stream}
% \caption{Legend (350 words max). Example legend text.}
% \label{fig:stream}
% \end{figure}

% \begin{table}[ht]
% \centering
% \begin{tabular}{|l|l|l|}
% \hline
% Condition & n & p \\
% \hline
% A & 5 & 0.1 \\
% \hline
% B & 10 & 0.01 \\
% \hline
% \end{tabular}
% \caption{\label{tab:example}Legend (350 words max). Example legend text.}
% \end{table}

% Figures and tables can be referenced in LaTeX using the ref command, e.g. Figure \ref{fig:stream} and Table \ref{tab:example}.

\clearpage
\section*{Supplementary Information}
\label{sec:supplement}
%\DeclareCaptionFormat{suppl}{Supplementary Figure \thefigure: #1#2#3}

\setcounter{figure}{0}
\setcounter{table}{0}
%\usepackage{caption}

% Customize the label name
\captionsetup[figure]{labelfont=bf, name=Supplementary Figure}
\captionsetup[table]{labelfont=bf, name=Supplementary Table}

\subsection*{Summary of Videos} 
Supplementary Table \ref{tab:experiment_data} provides an overview of the 2D+time macrophage videos used in the analysis, including experimental parameters and group annotations.
\begin{table}[h!]
	\centering
	\renewcommand{\arraystretch}{1.5}
	\begin{adjustbox}{max width=\textwidth}
		\begin{tabular}{c|c|c|c|c|c}
			\hline
			& \textbf{Video No.} & \textbf{X ($\mu m$)} & \textbf{Y ($\mu m$)} & \textbf{Time Step (min)} & \textbf{Time Range} \\
			\hline
			\multirow{5}{*}{M1} & 1 & $0.32$ & $0.32$ & 2.5 & 0.5--6.0 hpa\\
			& 2 & $0.32$ & $0.32$ & 2 & 0.6--6.0 hpa \\
			& 3 & $0.32$ & $0.32$ & 2.5 & 1.0--6.0 hpa\\
			& 4 & $0.32$ & $0.32$ & 1 & 2.0--6.0 hpa \\
			& 5 & $0.32$ & $0.32$ & 2.5 & 1.5--6.0 hpa\\
			\hline
			\multirow{2}{*}{M2} & 6 & $0.32$ & $0.32$ & 2.5 & 10.5--16.0 hpa\\
			& 7 & $0.32$ & $0.32$ & 2.5 & 10.5--16.0 hpa\\
			\hline
			\multirow{5}{*}{TP} & 8 & $0.32$ & $0.32$ & 2.5 & 6.0--10.5 hpa\\
			& 9 & $0.32$ & $0.32$ & 2.5 & 6.0--10.5 hpa\\
			& 10 & $0.32$ & $0.32$ & 2.5 & 6.0--10.5 hpa\\
			& 11 & $0.32$ & $0.32$ & 2.5 & 6.0--10.5 hpa\\
			& 12 & $0.32$ & $0.32$ & 2.5 & 6.0--10.5 hpa\\
			\hline
			\multirow{5}{*}{Non-M1} & 13 & $0.32$ & $0.32$ & 2.5 & 0.5--6.0 hpa\\
			& 14 & $0.32$ & $0.32$ & 2 & 0.6--6.0 hpa \\
			& 15 & $0.32$ & $0.32$ & 2.5 & 1.0--6.0 hpa\\
			& 16 & $0.32$ & $0.32$ & 1 & 2.0--6.0 hpa \\
			& 17 & $0.32$ & $0.32$ & 2.5 & 1.5--6.0 hpa\\
			\hline
			\multirow{3}{*}{M0} & 18 & $0.31$ & $0.31$ & 1.5 & 0--5.5 h \\
			& 19 & $0.31$ & $0.31$ & 1.5 & 0--5.5 h\\
			& 20 & $0.31$ & $0.31$ & 1.5 & 0--5.5 h\\
			\hline
		\end{tabular}
	\end{adjustbox}
	\caption{Overview of experimental video data, including spatial resolution, time intervals, recording durations, and annotations for different macrophage modes: M1, M2, Non-M1, and M0. TP denotes videos recorded during the transition period.}
	\label{tab:experiment_data}
\end{table}

\subsection*{Segmentation of Non-M1 Macrophages}
\begin{figure}[!ht]
	%\captionsetup{format=suppl, labelformat=empty} 
	\centering
	\includegraphics[scale=0.67]{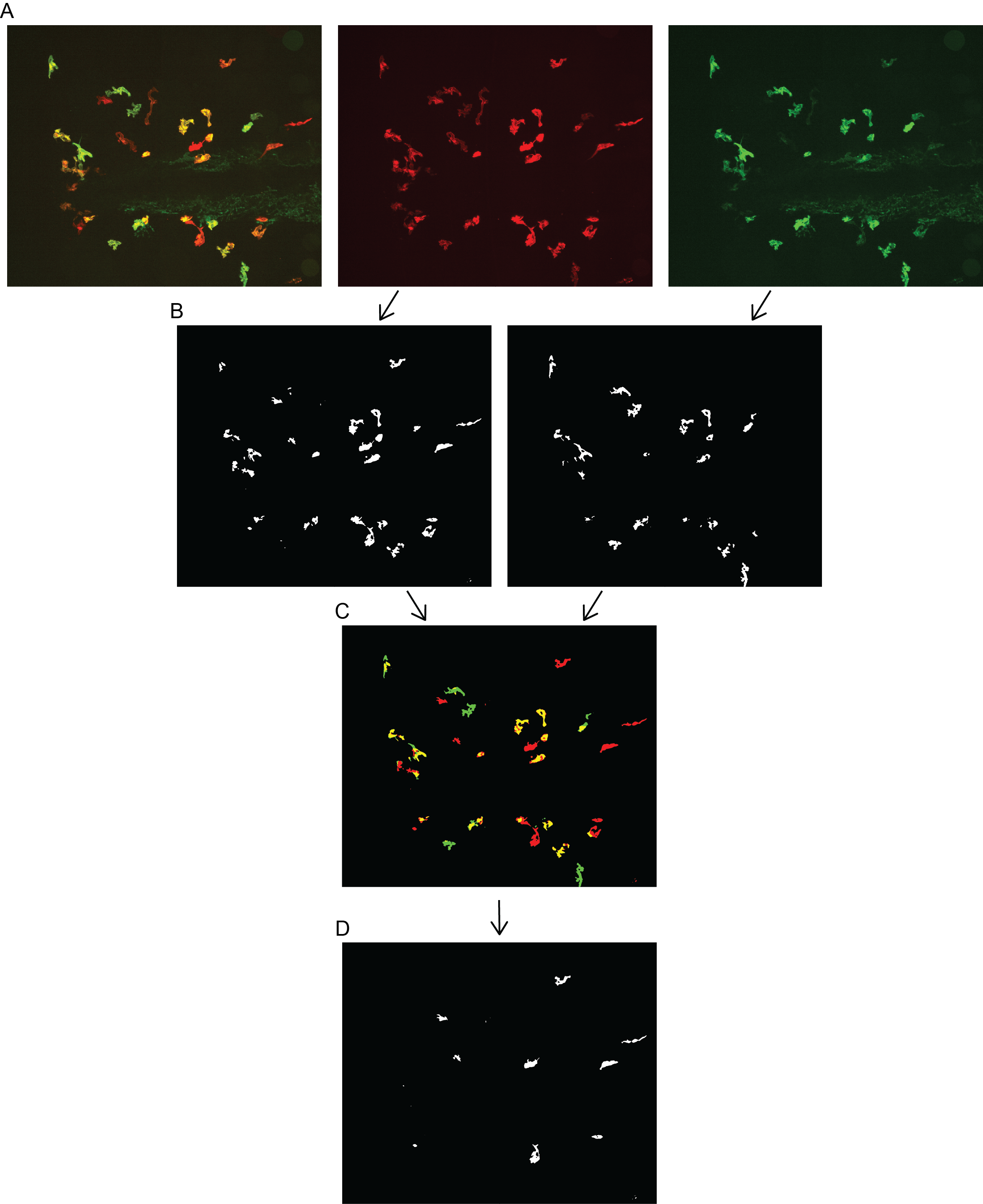}
	%   \vspace{3mm}
	\caption{A: Original data (Video 4) at 4.1 hpA, showing merged red and green channels (left), red channel only (middle), and green channel only (right). B: Segmented macrophages from the red (left) and green (right) channels. C: Green and red segmentations are overlaid; overlapping regions appear in yellow. D: Non-M1 macrophages identified by excluding any regions overlapping in both channels (yellow in C), even by a single pixel.}
	\label{Method:remove_overlap}
\end{figure}

Supplementary Figure \ref{Method:remove_overlap} provides a visual summary of the method for extracting Non-M1 macrophages.

\subsection*{Exclusion of Short and Unreliable Trajectories}
\begin{figure}[!ht]
	\centering
	\includegraphics[scale=0.67]{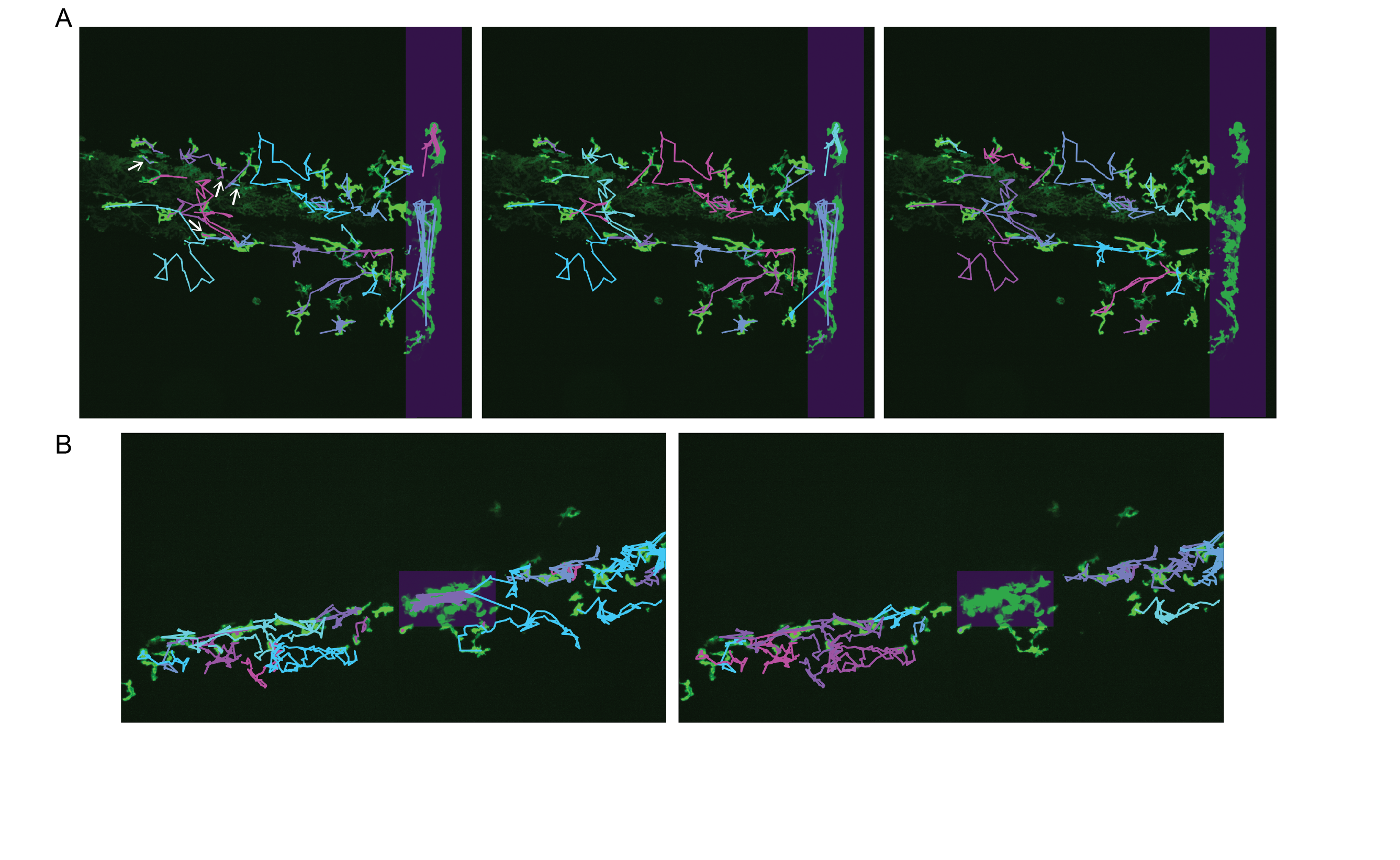}
	%   \vspace{3mm}
	\caption{A: The left panel shows all tracked trajectories in Video 6 at 13.6 hpA. Short trajectories (indicated by white arrows) containing fewer than 24 points, corresponding to a duration of 1 hour, were excluded, as shown in the middle panel. The right panel shows the final trajectories after excluding unreliable ones that remained within the MAZ (transparent violet region) for more than five consecutive time frames. B: Video 18 (M0) at 4.25 hours post-imaging, with the MAZ positioned in the central region. The left and right panels show the trajectories before and after applying the same steps of the trajectory exclusion as in A.}
	\label{Method:post_tra}
\end{figure}

Supplementary Figure \ref{Method:post_tra} illustrates examples of trajectory exclusion applied to M2 and M0 videos, including the removal of short trajectories and trajectories that exhibit limited presence within macrophage accumulation zones (MAZ), which are shown as manually defined transparent violet regions.

\subsection*{Description of the Morpho-Kinetic Features}
This section provides a brief overview of the morpho-kinetic features used in the study (Supplementary Table \ref{tab:features}), along with detailed mathematical definitions for each feature.

Some features were computed across all time frames within each trajectory. To summarize the temporal patterns of these features, we applied five statistical measures (mean, standard deviation, median, maximum, and minimum) to the time series of each feature within each trajectory. These are indicated by the prefixes mean\_, std\_, med\_, max\_, and min\_ preceding the feature names.

For kinetic features, we also analyzed smoothed versions of the trajectories to better capture their underlying trends while minimizing the effect of random parts. Features calculated from original and smoothed trajectories are labeled with the suffixes \_ori and \_smo, respectively.

\begin{table}[h!]
	\centering
	\caption{Summary of Features}
	\small % or \footnotesize for smaller text size
	\renewcommand{\arraystretch}{1.5} % Adjust the row height for better readability
	\begin{tabular}{>{\raggedright\arraybackslash}m{3cm}|>{\raggedright\arraybackslash}m{6cm}|>{\raggedright\arraybackslash}m{2cm}|>{\raggedright\arraybackslash}m{2cm}}
		\hline
		\textbf{Feature} & \textbf{Description} & \textbf{Statistics} & \textbf{Trajectory type} \\ \hline
		\textbf{Area (\text{area})} & The area of a macrophage from a trajectory at a time frame & mean\_, std\_, med\_, max\_, min\_ & None \\ \hline
		\textbf{Perimeter (\text{peri})} & The perimeter of a macrophage from a trajectory at a time frame & mean\_, std\_, med\_, max\_, min\_ & None \\ \hline
		\textbf{Circularity (\text{cir})} & The circularity of a macrophage from a trajectory at a time frame & mean\_, std\_, med\_, max\_, min\_ & None \\ \hline
		\textbf{Random Length Ratio (\text{r$_{\text{len}}$})} &  Ratio of the average length of random parts to the total length of the trajectory
		& None & None \\ \hline
		\textbf{Ratio of Random Segments (\text{r$_{\text{seg}}$})} & Ratio of line segments belonging to random parts relative to the total number of time frames
		& None & None \\ \hline
		\textbf{Meander Ratio (\text{M})} & Ratio of the total trajectory length to the displacement between the initial and final points & None & \_ori, \_smo \\ \hline
		\textbf{Tangent Projection Metrics (\text{TPM})} & Projection of trajectory tangent vectors onto the outward wound normal vector 
		& mean\_, std\_, med\_ max\_, min\_ & \_ori, \_smo \\ \hline
		\textbf{Positive TPM Ratio (\text{posT})} & Ratio of points with positive \texttt{TPM} values
		& None & \_ori, \_smo \\ \hline
		\textbf{Negative TPM Ratio (\text{negT})} & Ratio of points with negative \texttt{TPM} values
		& None & \_ori, \_smo \\ \hline
		\textbf{Velocity Projection Metrics (\text{VPM)}} & Projection of velocity vectors onto the outward wound normal vector 
		& mean\_, std\_, med\_, max\_, min\_ & \_ori, \_smo \\ \hline
		\textbf{Distance from Wound (\text{distW})} & Distances from the wound across all points of trajectories & mean\_, std\_, med\_, max\_, min\_ & \_ori, \_smo \\ \hline
		\textbf{Speed (\text{sp})} & Speeds across all points of trajectories & mean\_, std\_, med\_, max\_, min\_ & \_ori, \_smo \\ \hline
	\end{tabular}
	\label{tab:features}
\end{table}

\subsubsection*{Area, Perimeter, and Circularity}
Based on each approximate center of a macrophage, a connected region (i.e., a segmented macrophage) is identified using a flood fill algorithm on the thresholded image. For each detected connected region, the outer boundary is extracted using the marching squares algorithm. The perimeter (peri) is computed by summing the Euclidean distances between consecutive boundary points, while the area is calculated using the shoelace formula applied to the traced boundary coordinates. From the computed perimeter and area, the circularity is calculated as
\begin{equation} \label{cir}
	\text{cir} = 4\pi \frac{\text{area}}{\text{peri}^2}.
\end{equation}

\subsubsection*{Random Length Ratio}
Macrophage trajectories can be decomposed into random and directional parts. The definition and extraction process of the random components are detailed in \cite{lupi2025mathematical}. For each trajectory, the random length ratio is defined as

\begin{equation} \label{r_len}
	\text{r$_{\text{len}}$} = \frac{\langle L_{\text{rand}} \rangle}{L_{\text{total}}},
\end{equation}
where $\langle L_{\text{rand}} \rangle$ is the average length of random parts, and $L_{\text{total}}$ is the total length of original trajectory.
An example trajectory is shown in Supplementary Figure \ref{supple:random_parts}A, alongside its smoothed version, where random parts have been removed (details of the smoothing procedure are also provided in \cite{lupi2025mathematical}). 
The random parts, labeled as ``i'' and ``ii'' are illustrated as red segments. Their respective lengths are $78.43$ $\mu m$ and $114.78$ $\mu m$, with the total length of $673.41$ $\mu m$.

\subsubsection*{Ratio of Random Segments}
Using the extracted random parts, the ratio of random segments is defined as
\begin{equation} \label{r_seg}
	\texttt{r$_{\text{seg}}$} = \frac{N_{\text{rand}}}{N_{\text{total}}},
\end{equation}
where $N_{\text{rand}}$ is the number of line segments in random parts, and $N_{\text{total}}$ is the total number of segments in the trajectory.
For example, the random parts ``i'' and ``ii'' in Supplementary Figure \ref{supple:random_parts}A consist of 8 and 14 line segments, respectively. The total number of line segments in the trajectory is 53. Thus, $\text{r$_{\text{seg}}$} = \frac{8+14}{53}$.

\subsubsection*{Meander Ratio}
The meander ratio of a trajectory is defined as

\begin{equation} \label{meander}
	\text{M} = \frac{L}{d_{\text{start,end}}},
\end{equation}

where $L$ is the total length of the trajectory (either original or smoothed), and $d_{\text{start,end}}$ is the Euclidean distance between the start and end points of a trajectory.
The distance $d_{\text{start,end}}$ is the same for both the original and smoothed trajectories since the smoothing process preserves these points. However, the total trajectory length, $L$, differs between the original and smoothed versions. 
The meander ratio is computed for both cases and is denoted as \text{M$_{\text{ori}}$}(original trajectory) and \text{M$_{\text{smo}}$} (smoothed trajectory) in the main text.

\begin{figure}[!ht]
	\centering
	\includegraphics[width=\linewidth]{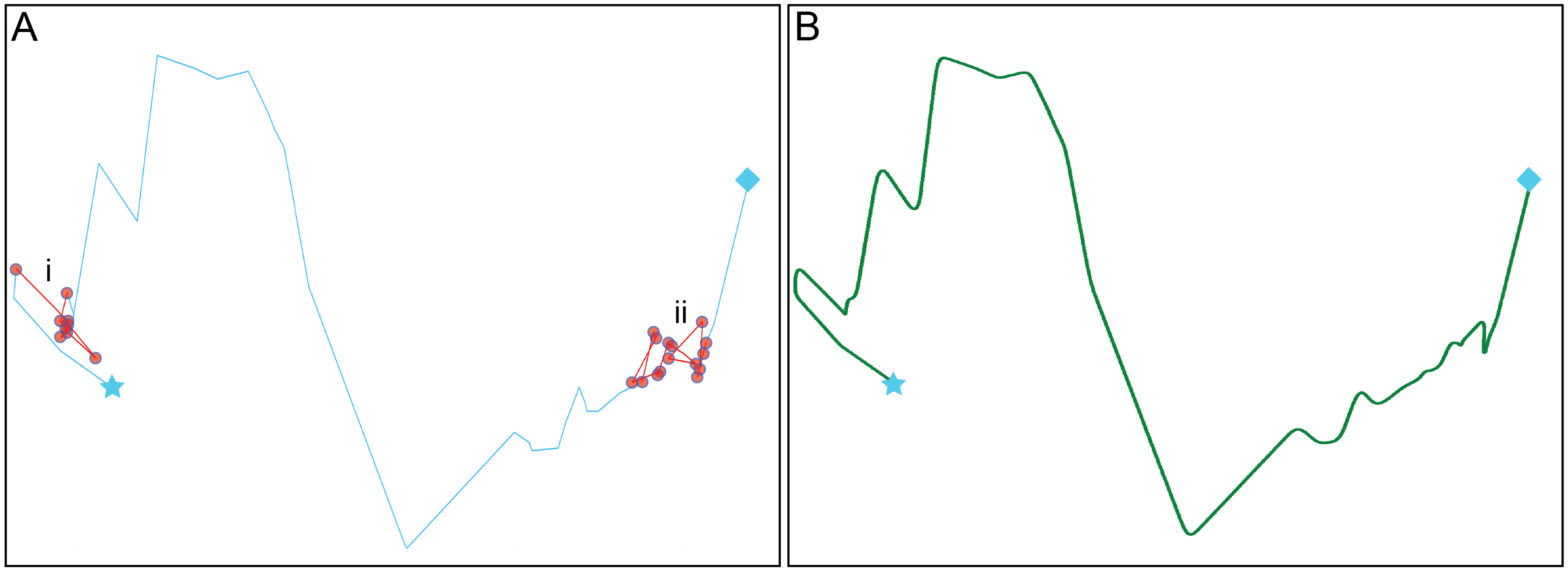}
	%   \vspace{3mm}
	\caption{A: Original extracted trajectory with two identified random parts, labeled ``i'' and ``ii''. B: Smoothed version of the trajectory in A. The start and end points are marked with a star and a diamond, respectively.}
	\label{supple:random_parts}
\end{figure}

\subsubsection*{Tangent Projection Metric}
This measure quantifies the directionality of macrophages, determining whether they are moving toward or away from the wound. For each point in a trajectory, the direction of the movement at the $i^{\text{th}}$ point is calculated using the tangent vector, which is defined as

\begin{equation} \label{tangent}
	\mathbf{T} = \frac{\mathbf{x}_{i+1} - \mathbf{x}_{i-1}}{\lVert \mathbf{x}_{i+1} - \mathbf{x}_{i-1}\rVert},
\end{equation}

where $\mathbf{x}_{i+1}$ represents the position vector at the $i+1^{th}$ point.
For both original and smoothed trajectories, the tangent projection metric (TPM) at the $i^{\text{th}}$ point is then defined as

\begin{equation} \label{TPM}
	\text{TPM} = \hat{\mathbf{N}}_w \cdot \mathbf{T},
\end{equation}

where $\hat{\mathbf{N}}_w$ is the outward normal vector of the wound (as shown in Supplementary Figure \ref{supple:T_VPM}). 
A positive $\text{TPM}$ indicates movement toward the wound, while a negative $\text{TPM}$ means movement away from the wound. 
%To summarize the directionality of each trajectory, we compute statistical measures, including the mean, standard deviation, median, maximum, and minimum TPM values. For example, the statistical values for an original trajectory are denoted as \texttt{mean\_TPM\_ori}, \texttt{std\_TPM\_ori}, \texttt{med\_TPM\_ori}, \texttt{max\_TPM\_ori}, and \texttt{min\_TPM\_ori}, respectively.

\subsubsection*{Positive and Negative TPM ratio}
To examine the directional tendency of trajectories, we calculate the ratio of points with positive or negative TPM values. These ratios are defined as
\begin{equation} \label{pos_neg_T}
	\begin{aligned}
		\text{posT} &= \frac{N_{\text{pos}}}{N_{\text{total}}}, \\
		\text{negT} &= \frac{N_{\text{neg}}}{N_{\text{total}}}, 
	\end{aligned}
\end{equation}
where $N_{\text{pos}}$ and $N_{\text{neg}}$ are the numbers of points with positive and negative TPM values, respectively. $N_{\text{total}}$ is the total number of trajectory points.
These ratios are calculated for both original and smoothed trajectories. For example, the positive TPM ratio for an original trajectory is denoted as \text{posT\_ori} in the main text.

\subsubsection*{Velocity Projection Metric}
This measure extends the concept of directionality by incorporating macrophage velocity. The velocity vector at the $i^{th}$ time frame is computed as

\begin{equation} \label{velocity}
	\mathbf{V} = \frac{\mathbf{x}_{i+1} - \mathbf{x}_{i-1}}{2\Delta t},
\end{equation}

where $\Delta t$ is the time interval between frames.
Similar to the computation of TPM, the velocity projection metric (VPM) is defined as

\begin{equation} \label{VPM}
	\text{VPM} = \hat{\mathbf{N}}_w \cdot \mathbf{V}.
\end{equation}

Note that Equation \eqref{velocity} applies to the original trajectories, while the velocity for the smoothed trajectories is computed using a different approach, as described in \cite{lupi2025mathematical}.
%As with TPM, VPM values are summarized for each trajectory using statistical measures such as the mean, standard deviation, median, maximum, and minimum. For example, for a smoothed trajectory, these values are denoted as \texttt{mean\_VPM\_smo}, \texttt{std\_VPM\_smo}, \texttt{med\_VPM\_smo}, \texttt{max\_VPM\_smo}, and \texttt{min\_VPM\_smo}, respectively.

\begin{figure}[!ht]
	\centering
	\includegraphics[scale=0.68]{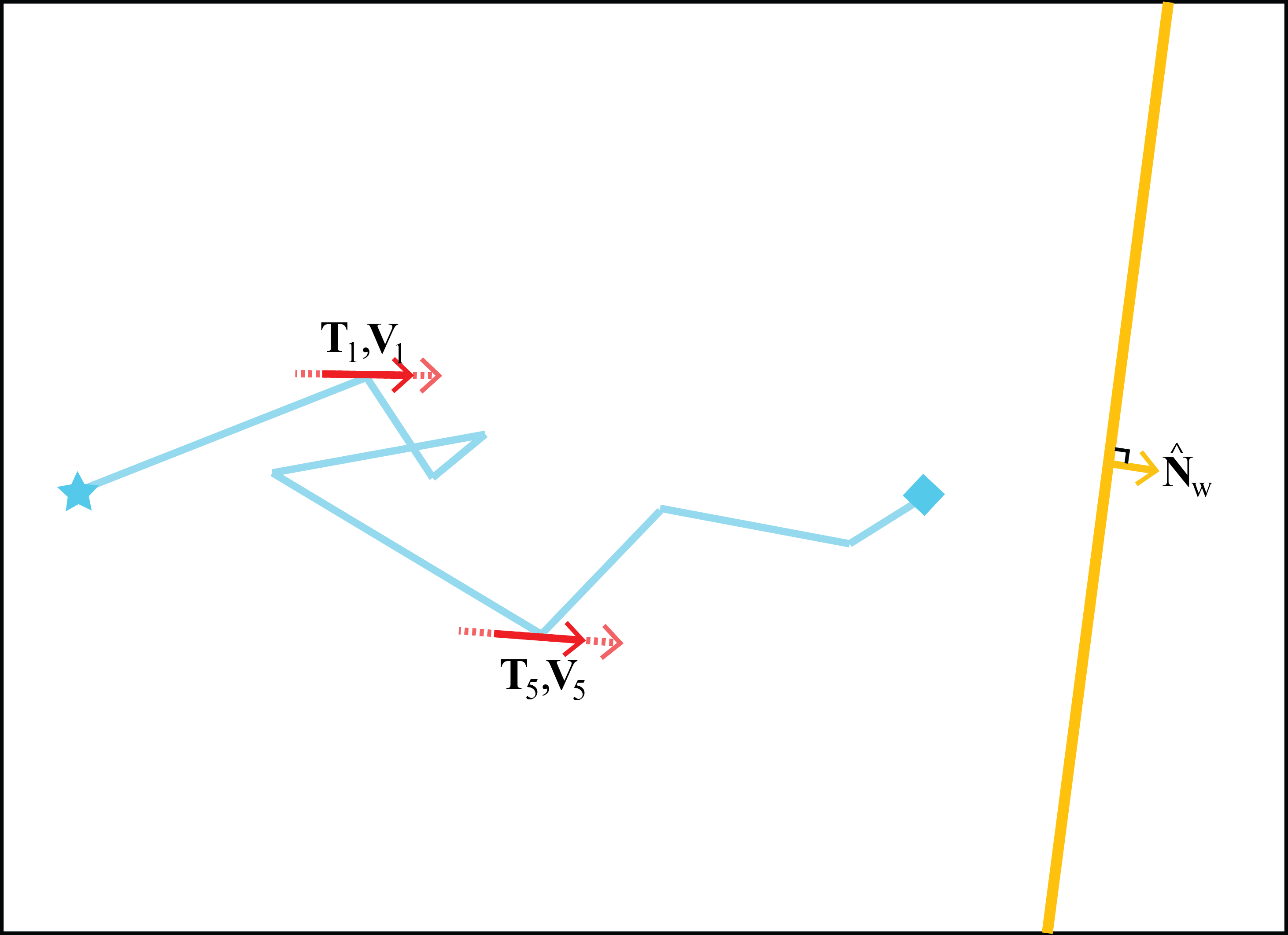}
	%   \vspace{3mm}
	\caption{$\mathbf{T}_{1}$, $\mathbf{V}_{1}$,  $\mathbf{T}_{5}$, and $\mathbf{V}_{5}$ illustrate the tangent (solid arrows) and velocity vectors (dashed arrows) at the first and fifth points of the trajectory.}
	\label{supple:T_VPM}
\end{figure}

\begin{figure}[!ht]
	\centering
	\includegraphics[scale=0.73]{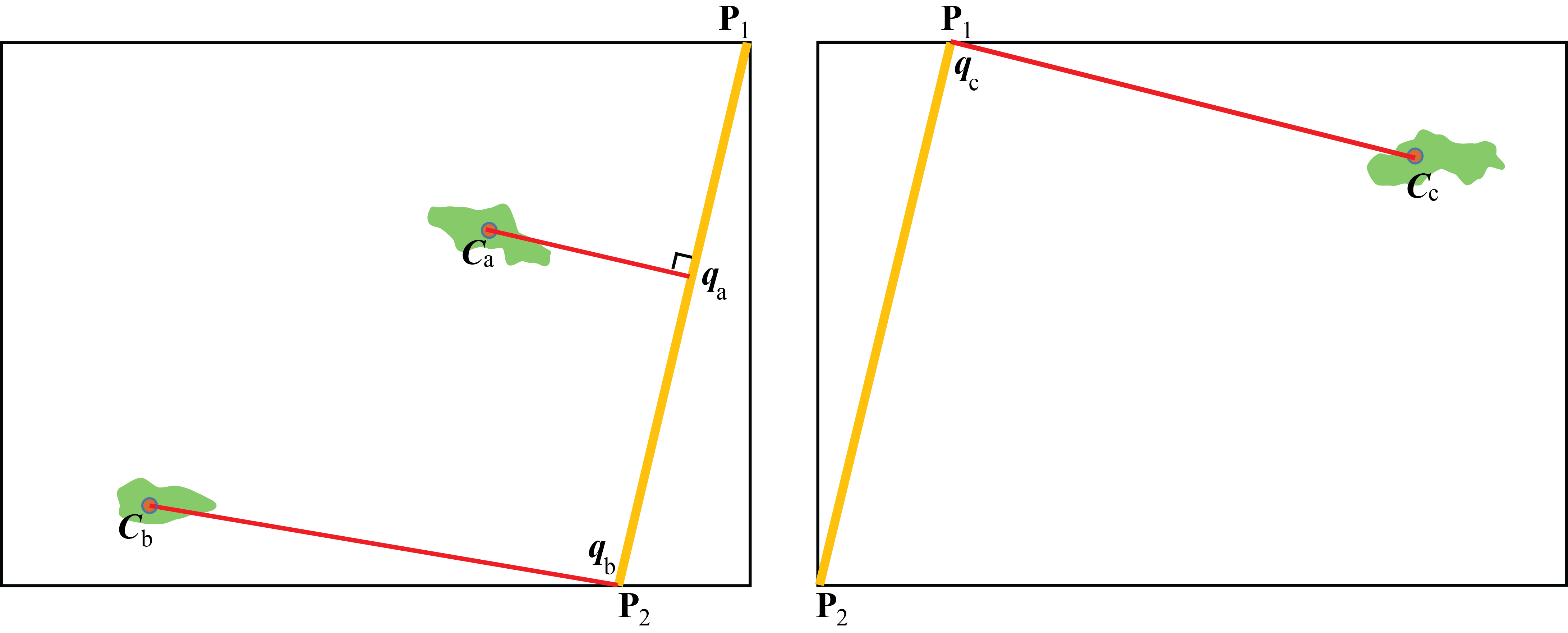}
	%   \vspace{3mm}
	\caption{Schematic illustration of the closest points on the wound site from individual macrophages. The wound is represented by two line segments with endpoints $\mathbf{P}_{1}$ and $\mathbf{P}_{2}$. The centers of three macrophages are denoted as $\textbf{\textit{C}}_{\text{a}}$, $\textbf{\textit{C}}_{\text{b}}$, and $\textbf{\textit{C}}_{\text{c}}$. The closest points corresponding to Cases a, b, and c are denoted by $\boldsymbol{q}_{a}$, $\boldsymbol{q}_{b}$, and $\boldsymbol{q}_{c}$, respectively.  }
	\label{supple:distW}
\end{figure}

\subsubsection*{Distance from Wound}
In this study, the wound is represented as a straight line defined by two manually selected endpoints, $\mathbf{P}_{1}$ and $\mathbf{P}_{2}$ (see Supplementary Figure~\ref{supple:distW}). To compute the distance from each trajectory point to the wound, we first identify the closest point on the wound line and then calculate the Euclidean distance.

In most cases, the closest point lies directly on the wound segment between $\mathbf{P}_{1}$ and $\mathbf{P}_{2}$, such that the vector from the trajectory point to this closest point is perpendicular to the wound line. This situation is illustrated by $\textbf{\textit{C}}_{\text{a}}$ in Supplementary Figure~\ref{supple:distW} (Case a). However, when the perpendicular projection falls outside the segment, such as in the cases of $\textbf{\textit{C}}_{\text{b}}$ and $\textbf{\textit{C}}_{\text{c}}$, the closest point is assigned to $\mathbf{P}{2}$ (Case b) or $\mathbf{P}_{1}$ (Case c), respectively.

\subsubsection*{Speed}
Speed is computed for every point along both the original and smoothed trajectories. For the original trajectories, the speed at the $i^{\text{th}}$ point is defined as the magnitude of the velocity vector, $\sqrt{\mathbf{V}\cdot\mathbf{V}}$. 
The speed for the smoothed trajectories is computed using the approach described in \cite{lupi2025mathematical}. 

\subsection*{Analytic Pipeline}
\begin{figure}[!ht]
	\centering
	\includegraphics[scale=0.7]{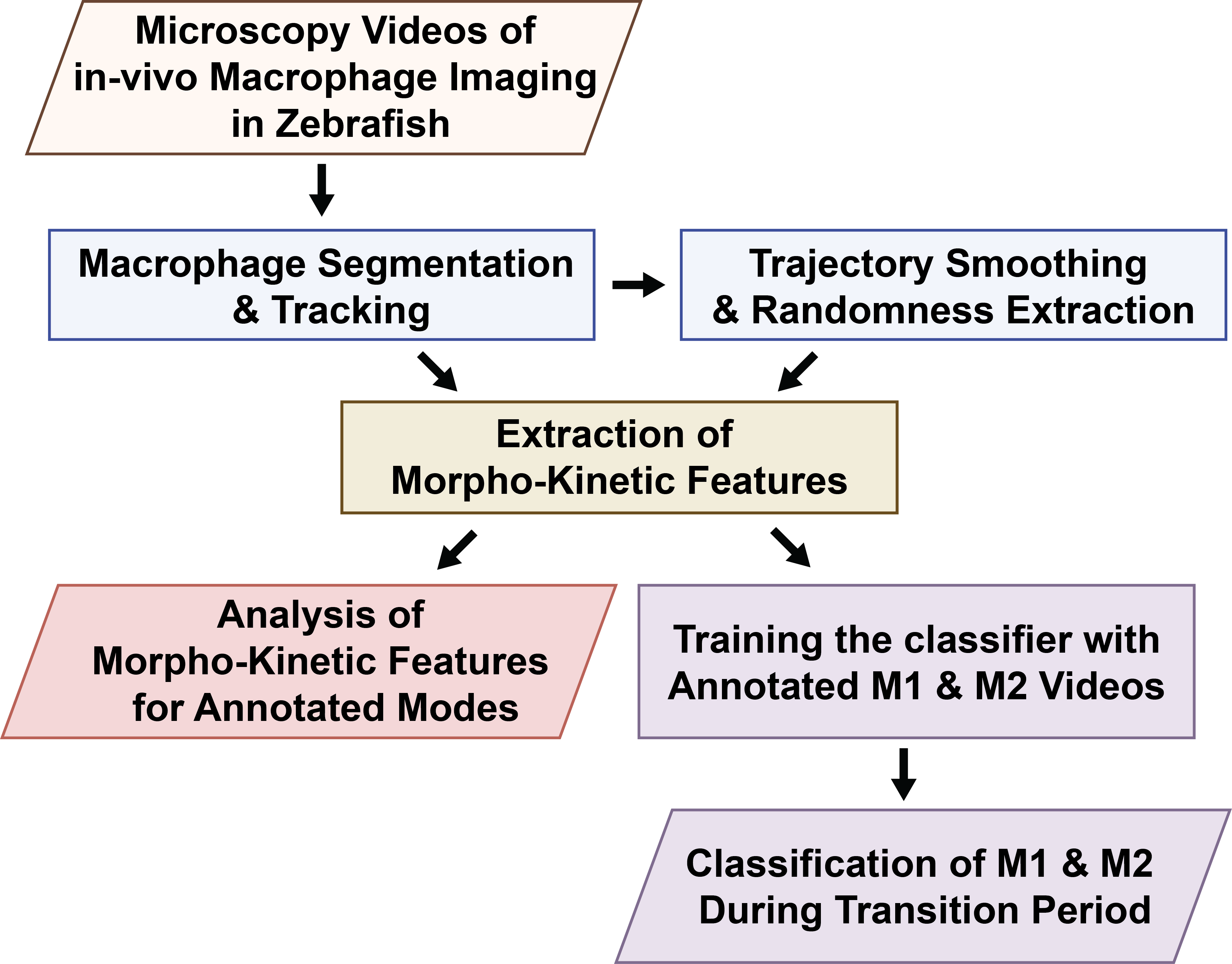}
	%   \vspace{3mm}
	\caption{Overview of the analytic pipeline, including image acquisition, data processing (segmentation, tracking, trajectory smoothing, randomness extraction), feature extraction, and macrophage classification and behavioral analysis. }
	\label{supple:pipeline}
\end{figure}

Supplementary Figure~\ref{supple:pipeline} outlines the overall analytic pipeline used in this study. The pipeline includes key stages such as macrophage segmentation and tracking, trajectory smoothing, extraction of random parts, quantitative feature extraction, and subsequent analysis to characterize macrophage behavior across different functional states and during the transition period.

%\newpage
%\section{Trajectory Point Distribution Over Time}
%Supplementary Figure~\ref{supple:avgpts} shows the temporal distribution of trajectory points, averaged across classified M1 and M2 macrophages in the transition period, to determine whether classified M1 and M2 trajectories follow a sequential pattern (with M1 occurring early and M2 occurring late) or coexist throughout the transition period (see also Discussion in the main text). The figure indicates that the latter is the case.

%\begin{figure}[!ht]
%	\captionsetup{format=suppl, labelformat=empty} 
%	\centering
%	\includegraphics[scale=0.5]{suppl_figure/avg_data_points_per_tra_bin_20.png}
%   \vspace{3mm}
%	\caption{Number of trajectory points per 20-minute time bin for each classified M1 and M2 macrophage trajectory. Values represent averages across all trajectories within each group.}
%	\label{supple:avgpts}
%\end{figure}

%\bibliographystyle{plainnat}
%\bibliography{suppl_reference}

\end{document}